\documentclass[3p]{elsarticle}

\AtBeginDocument{}
\usepackage{bm}
\usepackage[T1]{fontenc}
\usepackage{float}
\usepackage{tikz}
\usepackage{amssymb,amsbsy,amsmath,amsfonts,amssymb,amscd}
\usepackage{verbatim}
\usepackage{wrapfig}
\usepackage{algorithmic}
\usepackage{graphicx}
\usepackage{graphics}
\usepackage{stmaryrd}
\usepackage[small]{caption}
\usepackage{subcaption}
\usepackage{hyperref}
\usepackage{color}
\usepackage{amsthm}  
\usepackage{tabularx}
\usepackage{dcolumn}
\restylefloat{table}
\usepackage{booktabs}
\usepackage{xcolor,colortbl}
\usepackage{amssymb}
\usepackage{amsfonts}
\usepackage{graphicx}
\usepackage{parskip}

\newcommand{\norm}[1]{\left|\left|#1\right|\right|}

\journal{Computer Methods in Applied Mechanics and Engineering}

\begin{document}

\begin{frontmatter}

\title{Non-modal analysis of spectral element methods: Towards accurate and robust large-eddy simulations}

\author[add1]{Pablo Fernandez\corref{1}}
\ead{pablof@mit.edu}
\author[add2]{Rodrigo Moura\corref{2}}
\ead{moura@ita.br}
\author[add3]{Gianmarco Mengaldo\corref{3}}
\ead{mengaldo@caltech.edu}
\author[add1]{Jaime Peraire}
\ead{peraire@mit.edu}
\address[add1]{Department of Aeronautics and Astronautics, Massachusetts Institute of Technology, USA.}
\address[add2]{Instituto Tecnol{\' o}gico de Aeron{\' a}utica (ITA), Brazil.}
\address[add3]{Division of Engineering and Applied Science, California Institute of Technology, USA.}

\cortext[1]{Corresponding author}

\begin{abstract}
The use of high-fidelity flow simulations in conjunction with advanced numerical technologies, such as  spectral element methods for large-eddy simulation, is still very limited in industry. One of the main reasons behind this, is the lack of numerical robustness of spectral element methods for under-resolved simulations, than can lead to the failure of the computation or to inaccurate results, aspects that are critical in an industrial setting.  
In order to help address this issue, we introduce a \textit{non-modal} analysis technique that characterizes the diffusion properties of spectral element methods for linear convection-diffusion systems. 
While strictly speaking only valid for linear problems, the analysis is devised so that it can give critical insights on two questions: (i) Why do spectral element methods suffer from stability issues in under-resolved computations of nonlinear problems? And, (ii) why do they successfully predict under-resolved turbulent flows even without a subgrid-scale model? The answer to these two questions can in turn provide crucial guidelines to construct more robust and accurate schemes for complex under-resolved flows, commonly found in industrial applications. For illustration purposes, this analysis technique is applied to the hybridized discontinuous Galerkin methods as representatives of spectral element methods. The effect of the polynomial order, the upwinding parameter and the P\'eclet number on the so-called \textit{short-term diffusion} 
of the scheme are investigated. From a purely non-modal analysis point of view, polynomial orders between $2$ and $4$ with standard upwinding are well suited for under-resolved turbulence simulations. For lower polynomial orders, diffusion is introduced in scales that are much larger than the grid resolution. For higher polynomial orders, as well as for strong under/over-upwinding, robustness issues can be expected. The non-modal analysis results are then tested against under-resolved turbulence simulations of the Burgers, Euler and Navier-Stokes equations. While devised in the linear setting, our non-modal analysis succeeds to predict the behavior of the scheme in the nonlinear problems considered. The focus of this paper is on fluid dynamics, however, the results presented can be extended to other application areas characterized by under-resolved scales.
\end{abstract}


\end{frontmatter}

\section{\label{s:introduction}Introduction}

The use of computational fluid dynamics (CFD) in industry is severely limited by the inability to accurately and reliably predict complex turbulent flows. This is partly due to the current numerical technologies adopted by industry practitioners, that still rely on steady-tailored techniques, in conjunction with low-order numerical methods. In fact, the majority of the numerical codes adopted in fluid dynamics have first or second order spatial accuracy and are based on Reynolds-Averaged Navier-Stokes (RANS) equations or, more recently, detached-eddy simulation (DES). The use of high-fidelity computer-aided design is still very limited, with large-eddy simulation (LES) largely confined in the research and development branches of industries, or in academia. However, with the increase in computing power, LES is becoming a feasible technique to underpin the complexity of challenging industrial flows at high-Reynolds numbers, and spectral element methods (also referred to as spectral/$hp$ methods, briefly SEM) are a competitive candidate to improve the performance of the overall computer-aided workflow \cite{moxey2017towards}. In fact, the adoption of SEM in the context of LES, including the use of continuous Galerkin (CG) methods \cite{Karamanos:2000,Kirby:2003}, standard discontinuous Galerkin (DG) methods \cite{Beck:14,Frere:15,Gassner:13,Mengaldo:2015-Triple,Murman:16,Renac:15,Uranga:11,Wiart:15}, hybridized DG methods \cite{Fernandez:16a,Fernandez:17a}, spectral difference (SD) methods \cite{Liang:2009,Parsani:2010} and flux reconstruction (FR) methods \cite{Park:2017,Vermeire:2014}, is emerging as a promising approach to solve complex turbulent flows. First, they allow for high-order discretizations on complex geometries and unstructured meshes. This is critical to accurately propagate small-scale, small-magnitude features, such as in transitional and turbulent flows, over the complex three-dimensional geometries commonly encountered in industrial applications. Second, SEM are highly tailored to emerging computing architectures, including graphics processing units (GPUs) and many-core architectures, due to their high flop-to-communication ratio. The use of these methods for LES is being further encouraged by successful numerical predictions (see references before).

Yet, a critical step needs to be overcome in order to take advantage of the favorable properties of SEM for LES, that is the lack of robustness of these methods for under-resolved simulations \cite{mengaldo2015dealiasing,mengaldo2015discontinuous,winters2017comparative}. 
From this perspective, the diffusion characteristics of the discretization scheme, which are still not well understood for SEM, play a critical role in the robustness, as well as in the accuracy, of under-resolved computations. In particular, it is critical to understand the numerical diffusion introduced by the scheme, including the scales affected by numerical diffusion and the relationship between the amount of numerical diffusion and the `level' of under-resolution in the simulation. These points will help provide an answer on why certain SEM computations suffer from numerical stability issues, while others are robust and successfully predict under-resolved turbulent flows even without a subgrid-scale model \cite{Fernandez:2018:DGfluid,moura2017eddy,moura2017setting}. As customary, we use the term \textit{under-resolved} to refer to simulations in which the exact solution contains scales that are smaller than the Nyquist wavenumber of the grid (the so-called subgrid scales) and thus cannot be captured with the grid resolution. 

For numerical schemes with more than one degree of freedom (DOF) per computational cell, such as in high-order SEM, several ways of investigating the diffusion characteristics of the scheme are possible. The most widely used technique is the eigensolution analysis; which has been succesfully applied to CG \cite{Moura:16b}, standard DG \cite{Hu:1999,Hu:2002,Mengaldo:ComputersFluids:2017,Moura:15a}, hybridized DG \cite{Moura:18a} and FR \cite{Mengaldo:eigenFR:2018a} methods. 
Eigenanalyses address the diffusion and dispersion characteristics, in wavenumber space, of the discretization of linear propagation-type problems, such as the linear convection or convection-diffusion equation in one dimension. In the SEM context, Fourier modes $\exp(i \kappa x)$, where $\kappa$ denotes wavenumber, are in general not eigenmodes of the discretization, and are therefore given by the contribution of several of these eigenmodes. In eigenanalysis, typically all but one of the eigenmodes are dismissed as secondary (or unphysical), and the focus is placed on the so-called primary (or physical) eigenmode \cite{Moura:15a,Moura:16b,Moura:18a}. Although the primary mode is the one that more clearly represents well-resolved Fourier modes (i.e.\ well-resolved wavenumbers), the secondary eigenmodes can strongly influence the solution characteristics for wavenumbers near the Nyquist wavenumber. As a consequence, the primary eigenmodes accurately characterize the behavior of the scheme in well-resolved simulations, but may fail in under-resolved computations in which scales near the Nyquist wavenumber play an important role. Alternatively, because eigenmodes are decoupled in linear problems, one can consider only the least dissipated one (which may not be the primary one) to assess the long-term dynamics of the problem at hand. In nonlinear settings, all eigenmodes are potentially coupled, and one can only expect that assessing their combined effect will lead to a better understanding of the numerical characteristics relevant to under-resolved turbulence computations.

As an alternative and complementary approach to eigenanalysis, we are interested in the actual (i.e.\ non-modal) short-term dynamics of the discretization of the linear convection-diffusion equation. This is motivated by the idea that nonlinear dynamical systems behave similarly to its linearized version during a short period of time, and it is some form of the short-term behavior of the linearized system that is likely to be most informative about the nonlinear dynamics. To this end, we introduce a \textit{new} analysis framework and refer to it as \textit{non-modal analysis} as it resembles non-modal stability theory\footnote{Non-modal stability theory studies the transient growth of non-modal disturbances in linear dynamical systems (non-modal in the sense that they are not eigenmodes) and was a major breakthrough to characterize nonlinear instabilities by analyzing the short-term behavior of the linearized dynamics.} \cite{Schmid:2007,Trefethen:1993,Trefethen:1997}. Our non-modal analysis does not require Fourier modes to be eigenmodes of the discretization and reconciles with eigenanalysis whenever they actually are. In particular, it is informative of the behavior of the numerical scheme for time instants immediately after an initial condition is prescribed. For this reason, we take the liberty of using the term `short-term' whenever referring to the behavior of the numerical discretization described by the non-modal analysis framework proposed in this paper. We note that the numerical solution at any time can be thought of as an initial condition for the remaining of the simulation, and this interpretation is particularly useful when employing linear techniques to analyze complex nonlinear systems, such as turbulent flows at high-Reynolds numbers. From this perspective, we assess how the non-modal analysis applies to nonlinear problems, including Burgers, Euler and Navier-Stokes equations, and we discuss guidelines on how to construct more robust numerical schemes, from the insights obtained from the analysis. 
While the terms diffusion and dissipation are commonly used interchangeably in the literature, in this paper we reserve the latter for situations in which it is some form of `energy' that is affected, such as kinetic energy in the nonlinear examples considered.

The remainder of the paper is organized as follows. In Section \ref{s:nonModal_HDG}, we introduce the non-modal analysis framework and apply it to the hybridized discontinuous Galerkin methods. The short-term diffusion characteristics of hybridized DG methods are investigated in this section. In Section \ref{s:applicationNonLinear}, we assess how non-modal analysis results extend to the nonlinear setting. To that end, we compare non-modal analysis with numerical results for the Burgers, Euler and Navier-Stokes equations. A discussion on how to devise superior schemes using insights from non-modal analysis is presented in Section \ref{s:discussion}. We conclude the paper with some remarks in Section \ref{s:conclusions}.

\section{\label{s:nonModal_HDG}Non-modal analysis for hybridized DG}

We illustrate the non-modal analysis framework by applying it to the hybridized DG methods \cite{Nguyen:15,Fernandez:17a}, a particular instance of high-order SEM. More precisely, the hybridized DG methods are a class of discontinuous Galerkin methods that generalizes the Hybridizable DG (HDG) \cite{Cockburn:09a,Nguyen:12}, Embedded DG (EDG) \cite{Cockburn:09a,Cockburn:09b} and Interior Embedded DG (IEDG) \cite{Fernandez:16a} methods. This family of schemes is becoming increasingly popular for fluid mechanics \cite{Dahm:2014,Fernandez:17a,Giorgiani:2014,Lehrenfeld:2016,Rhebergen:2012,Schutz:2013,Ueckermann:2016,Woopen:2014}, solid mechanics \cite{Sheldon:2016,Terrana:2018} and electromagnetism \cite{Chen:2015,Christophe:2018,Feng:2016,Li:2017,VidalCodina:2018,Yoo:2016} as it allows (for moderately high accuracy orders) for more computationally efficient implementations than standard DG methods. The analysis herein can be readily extended to other high-order SEM, both continuous and discontinuous, including CG, standard DG, SD, and FR methods.

We note that hybridized and standard DG methods are equivalent for the linear convection equation (see \ref{s:connectionStHyDG}), and therefore the hybridized DG non-modal analysis for pure convection also applies to standard DG. Similarly, the results for pure convection carry over to certain types of FR schemes (see \cite{DeGrazia:2014,Mengaldo:2016,Vincent:2011} for the connections between DG and FR methods).

\subsection{Spatial discretization of the linear convection-diffusion equation}

We consider the linear convection-diffusion equation with constant coefficients in a one-dimensional domain $\Omega = (-\infty , \infty)$, given by
\begin{subequations}
\label{e:lcd}
\begin{alignat}{2}
& \frac{\partial u}{\partial t} + a \frac{\partial u}{\partial x} = \nu \frac{\partial ^2 u}{\partial x^2} , \qquad & t > 0 , \\
& u = u_0 , \qquad & t = 0 , 
\end{alignat}
\end{subequations}
where $a$ is the convection velocity, $\nu \geq 0$ is the diffusion coefficient, and $u_0 \in \mathcal{C}^2(\mathbb{R} ; \mathbb{C})$ is a twice continuously differentiable (possibly complex-valued) initial condition. In order to discretize Eq. \eqref{e:lcd} in space by hybridized DG methods, we first rewrite it in the following conservation form
\begin{subequations}
\label{e:lcd2}
\begin{alignat}{2}
\label{e:lcd2_1}
& \frac{\partial u}{\partial t} + \frac{\partial f}{\partial x} = 0 , \qquad & t > 0 \mbox{ ,} \\
\label{e:lcd2_2}
& q = \frac{\partial u}{\partial x} , \qquad & t \geq 0 \mbox{ ,} \\
\label{e:lcd2_3}
& u = u_0 , \qquad & t = 0 \mbox{ ,} 
\end{alignat}
\end{subequations}
where $q$ is the auxiliary gradient variable and $f(u,q) = a \, u - \nu \, q$ is the flux function. After $\Omega$ is partitioned into uniform non-overlapping elements $\Omega_e$ of size $h$, the numerical solution and its gradient in a given element $\Omega_e$ are approximated by polynomial expansions of the form
\begin{equation} \label{eq:exp}
u_h|_{\Omega_e} = \sum_{j=0}^{P}{\tilde{u}_{h,j}(t) \, \phi_j(\xi(x))} \mbox{ ,} \qquad \qquad q_h|_{\Omega_e} = \sum_{j=0}^{P}{\tilde{q}_{h,j}(t) \, \phi_j(\xi(x))} \mbox{ .} 
\end{equation}
where $\phi_j$ are polynomial basis functions of degree up to $P$, defined in the reference domain $\Omega_{\textnormal{ref}} = [-1,1]$. A linear mapping relation is assumed between the physical coordinate $x$ in element $\Omega_e$ and the coordinate $\xi = \xi(x) \in \Omega_{\textnormal{ref}}$. Multiplying Equations \eqref{e:lcd2_1}$-$\eqref{e:lcd2_2} by $\phi_i$, integrating over $\Omega_e$, and applying integration by parts leads to
\begin{subequations}
\label{eq:disc}
\begin{alignat}{2}
\label{eq:discU}
& \frac{h}{2} \int_{\Omega_{\textnormal{ref}}}{\frac{\partial u_h}{\partial t} \phi_i} + \left( \widehat{f}_h \phi_i \right)^{\oplus}_{\ominus} = \int_{\Omega_{\textnormal{ref}}}{f \frac{\partial \phi_i}{\partial \xi}} \mbox{ ,} \\
\label{eq:discG}
& \frac{h}{2} \int_{\Omega_{\textnormal{ref}}}{q_h \phi_i} + \int_{\Omega_{\textnormal{ref}}}{u_h \frac{\partial \phi_i}{\partial \xi}} = \big( \widehat{u}_h \phi_i \big)^\oplus_\ominus \mbox{ ,} 
\end{alignat}
\end{subequations}
where the symbols $\ominus$ and $\oplus$ denote the left and right boundaries of $\Omega_e$, respectively. As customary in the context of DG methods, expressions \eqref{eq:exp} are inserted into \eqref{eq:disc}; which are then required to hold for $i = 0, \dots, P$. Note that we have introduced the interface quantities $\widehat{f}_h$ and $\widehat{u}_h$. The former is the so-called interelement flux, interface flux or numerical flux, and appears in standard DG methods as well. The latter is particular of hybridized DG methods and is an approximation for the solution $u$ on the element faces that takes the same value on the two elements neighboring the considered interface.

In order to complete the definition of the hybridized DG scheme, it remains to define the numerical flux $\widehat{f}_h$ and enforce its continuity from the left ($L$) to the right ($R$) elements sharing the interface. For convection-diffusion problems, the numerical fluxes are usually defined as \cite{Nguyen:09,Peraire:10}
\begin{subequations}
\label{eq:flux_}
\begin{alignat}{2}
\label{eq:fluxmi}
\widehat{f}_{h,\ominus} = f(\widehat{u}_{h,\ominus}, q_{h,\ominus}) - \sigma (u_{h,\ominus} - \widehat{u}_{h,\ominus}) \mbox{ ,} \\
\label{eq:fluxpl}
\widehat{f}_{h,\oplus} = f(\widehat{u}_{h,\oplus}, q_{h,\oplus}) + \sigma (u_{h,\oplus} - \widehat{u}_{h,\oplus}) \mbox{ ,}
\end{alignat}
\end{subequations}
where $\sigma = \beta \, |a|$ is the stabilization constant and $\beta \geq 0$ the so-called upwinding parameter. The case $\beta = 1$ corresponds to standard upwinding, whereas $\beta = 0$ corresponds to centered convective fluxes. Also, $0 < \beta \ll 1$ and $\beta \gg 1$ will be referred to as strong under- and over-upwinding, respectively. Note that no explicit stabilization is used for the diffusive term. Since hybridized DG methods have some form of built-in stabilization for second-order operators \cite{Fernandez:PhD:2018}, this choice of $\sigma$ is customary for high Reynolds number flows \cite{Fernandez:17a,Nguyen:2011,Peraire:10} and has been adopted here for consistency with the literature. The flux continuity condition is then given by
\begin{equation} \label{eq:fluxcont}
\widehat{f}_{h,\oplus}^L = \widehat{f}_{h,\ominus}^R \mbox{ .}
\end{equation}
We note that Eq. \eqref{eq:fluxcont} ensures local conservation regardless of the chosen numerical flux formula. Also, for pure convection and our choice of numerical fluxes, it follows that $\widehat{u}_h = (u_{h,\oplus}^L + u_{h,\ominus}^R) / 2$ and, furthermore, hybridized and standard DG methods lead to the same numerical solution (see \ref{s:connectionStHyDG}). This does not hold, however, when diffusion is taken into account. In this case $\widehat{u}_h$ is only given implicitly from the flux continuity at interfaces, namely,
\begin{equation} \label{eq:fluxcont2}
a \widehat{u}_h - \nu q_{h,\oplus}^L + \sigma ( u_{h,\oplus}^L - \widehat{u}_{h} ) = a \widehat{u}_h - \nu q_{h,\ominus}^R - \sigma ( u_{h,\ominus}^R - \widehat{u}_h ) \mbox{ ,}
\end{equation}
where $q_{h,\oplus}^L$ and $q_{h,\ominus}^R$ in turn depend on the values of $\widehat{u}_h$ at two other interfaces via \eqref{eq:discG}. 

To simplify the analysis, we rewrite the hybridized DG discretization in matrix notation. To that end, we note that
\begin{subequations}
\begin{alignat}{2}
\label{eq:minusUG}
u_{h,\ominus} = \sum_{j=0}^{P}{\tilde{u}_{h,j} \, \phi_j(-1)} \mbox{ ,} \quad q_{h,\ominus} = \sum_{j=0}^{P}{\tilde{q}_{h,j} \, \phi_j(-1)} \mbox{ ,} \\
\label{eq:plusUG}
u_{h,\oplus} = \sum_{j=0}^{P}{\tilde{u}_{h,j} \, \phi_j(+1)} \mbox{ ,} \quad q_{h,\oplus} = \sum_{j=0}^{P}{\tilde{q}_{h,j} \, \phi_j(+1)} \mbox{ ,}
\end{alignat}
\end{subequations}
and introduce the vectors $\tilde{u}_h = \{ \tilde{u}_0, \dots, \tilde{u}_P \}^T$, $\tilde{q} = \{ \tilde{q}_0, \dots, \tilde{q}_P \}^T$, $\tilde{\phi}_\oplus = \{ \tilde{\phi}_0(+1), \dots, \tilde{\phi}_P(+1) \}^T$ and $\tilde{\phi}_\ominus = \{ \tilde{\phi}_0(-1), \dots, \tilde{\phi}_P(-1) \}^T$. The flux continuity condition \eqref{eq:fluxcont} then can be expressed as
\begin{equation} \label{eq:III}
\widehat{u}_h = \frac{1}{2} \left( \tilde{\phi}^T_{\oplus} \tilde{u}_h^L + \tilde{\phi}^T_\ominus \tilde{u}_h^R \right) + \frac{\nu}{2 \sigma} \left( \tilde{\phi}^T_\ominus \tilde{q}_h^R - \tilde{\phi}^T_\oplus \tilde{q}_h^L \right) \mbox{ .}
\end{equation}
Likewise, the auxiliary equation \eqref{eq:discG} can be written as
\begin{equation} \label{eq:II}
\frac{h}{2} M \tilde{q}_h + D \tilde{u}_h = \tilde{\phi}_\oplus \widehat{u}_{h,\oplus} - \tilde{\phi}_\ominus \widehat{u}_{h,\ominus} \mbox{ ,}
\end{equation}
where $M$ and $D$ are the mass and convection matrices defined as
\begin{equation} \label{eq:matsMD}
M_{ij} = \int_{\Omega_{\textnormal{ref}}} \phi_i \, \phi_j \mbox{ ,} \qquad \qquad D_{ij} = \int_{\Omega_{\textnormal{ref}}} \frac{\partial \phi_i}{\partial \xi} \, \phi_j \mbox{ .}
\end{equation}
Finally, Eq. \eqref{eq:discU} becomes
\begin{equation} \label{eq:I}
\frac{h}{2} M \frac{d \tilde{u}_h}{d t} + \tilde{\phi}_\oplus \widehat{f}_{h,\oplus} - \tilde{\phi}_\ominus \widehat{f}_{h,\ominus} = a D \tilde{u}_h - \nu D \tilde{q}_h \mbox{ ,}
\end{equation}
with
\begin{subequations}
\label{eq:fluxI}
\begin{equation} \label{eq:fluxImi}
\widehat{f}_{h,\ominus} = a \widehat{u}_{h,\ominus} - \nu \tilde{\phi}^T_\ominus \tilde{q}_h - \sigma ( \tilde{\phi}^T_\ominus \tilde{u}_h - \widehat{u}_{h,\ominus} ) \mbox{ .}
\end{equation}
\begin{equation} \label{eq:fluxIpl}
\widehat{f}_{h,\oplus} = a \widehat{u}_{h,\oplus} - \nu \tilde{\phi}^T_\oplus \tilde{q}_h + \sigma ( \tilde{\phi}^T_\oplus \tilde{u}_h - \widehat{u}_{h,\oplus} ) \mbox{ ,}
\end{equation}
\end{subequations}
Note that \eqref{eq:III} is a scalar equation written from the point of view of a given interface, whereas \eqref{eq:II} and \eqref{eq:I} are vector equations written from the viewpoint of an arbitrary element $\Omega_e$ of size $h$.

At this point, it is convenient to eliminate $\tilde{q}_h$ from the formulation and work with the variables $\tilde{u}_h$ and $\widehat{u}_h$ only. To this end, we use \eqref{eq:II} to obtain $\tilde{q}_h$ as a function of $\tilde{u}_h$ and $\widehat{u}_h$, and then substitute the resulting expression into \eqref{eq:III} and \eqref{eq:I}. The former substitution leads, after some algebra, to
\begin{equation} \label{eq:star}
\left( \bar{\sigma} + \frac{m_\ominus^\ominus}{\mbox{Pe}} + \frac{m_\oplus^\oplus}{\mbox{Pe}} \right) \widehat{u}_h - \frac{m_\oplus^\ominus}{\mbox{Pe}} \widehat{u}_{h,\ominus}^L - \frac{m_\ominus^\oplus}{\mbox{Pe}} \widehat{u}_{h,\oplus}^R = \tilde{\phi}^T_\oplus B_\oplus \tilde{u}_h^L + \tilde{\phi}^T_\ominus B_\ominus \tilde{u}_h^R \mbox{ ,}
\end{equation}
where $\bar{\sigma} = \sigma / |a|$ is a non-dimensional stabilization parameter and $\mbox{Pe}$ denotes the cell P\'eclet number $\mbox{Pe} = |a| h / \nu$. Note $\bar{\sigma} = \beta$ for our choice of stabilization. Moreover, the following scalar constants `$m$'
\begin{equation} \label{eq:emes}
m_{\ominus}^{\ominus} = \tilde{\phi}_{\ominus}^T M^{-1} \tilde{\phi}_{\ominus} \mbox{ ,} \quad
m_{\oplus}^{\ominus} = \tilde{\phi}_{\oplus}^T M^{-1} \tilde{\phi}_{\ominus} \mbox{ ,} \quad
m_{\ominus}^{\oplus} = \tilde{\phi}_{\ominus}^T M^{-1} \tilde{\phi}_{\oplus} \mbox{ ,} \quad
m_{\oplus}^{\oplus} = \tilde{\phi}_{\oplus}^T M^{-1} \tilde{\phi}_{\oplus} \mbox{ .}
\end{equation}
and matrices
\begin{equation}
B_{\ominus} = \bigg( \frac{\bar{\sigma}}{2} \, I - \frac{M^{-1} D}{\textnormal{Pe}} \bigg) \mbox{ ,} \qquad \qquad B_{\oplus} = \bigg( \frac{\bar{\sigma}}{2} \, I + \frac{M^{-1} D}{\textnormal{Pe}} \bigg) \mbox{ ,}
\end{equation}
have been introduced in \eqref{eq:star}. Note that Eq. \eqref{eq:star} links the solution vectors $\tilde{u}_h$ on two adjacent elements ($\Omega_L$ and $\Omega_R$) with the three interface variables $\widehat{u}_h$ corresponding to the boundaries of these elements.

The second substitution, namely inserting $\tilde{q}_h$ from \eqref{eq:II} into \eqref{eq:I}, and using also Equation \eqref{eq:fluxI} for the numerical fluxes, yields
\begin{equation} \label{eq:aster}
\frac{h}{2 a} M \frac{d \tilde{u}_h}{d t} + A \tilde{u}_h = A_\ominus \tilde{\phi}_\ominus \widehat{u}_{h,\ominus} + A_\oplus \tilde{\phi}_\oplus \widehat{u}_{h,\oplus} \mbox{ ,}
\end{equation}
where
\begin{subequations}
\begin{equation}
\label{eq:matA1}
A = \bar{\sigma} \, \big( \Phi_{\ominus}^{\ominus} + \Phi_{\oplus}^{\oplus} \big) + \bigg( \frac{2N}{\textnormal{Pe}} - I \bigg) \, D \mbox{ ,}
\end{equation} 
\begin{equation}
\label{eq:matA2}
A_{\ominus} = (\bar{\sigma} + 1) \, I - \frac{2 N}{\textnormal{Pe}} \mbox{ ,} \qquad \qquad A_{\oplus} = (\bar{\sigma} - 1) \, I + \frac{2 N}{\textnormal{Pe}} \mbox{ ,}
\end{equation} 
\end{subequations}
and
\begin{subequations}
\begin{equation}
\label{eq:matsPhi}
\Phi_{\ominus}^{\ominus} = \tilde{\phi}_{\ominus} \tilde{\phi}_{\ominus}^T \mbox{ ,} \qquad \qquad \Phi_{\oplus}^{\oplus} = \tilde{\phi}_{\oplus} \tilde{\phi}_{\oplus}^T \mbox{ ,}
\end{equation} 
\begin{equation}
\label{eq:matN}
N = \big( \Phi_{\oplus}^{\oplus} - \Phi_{\ominus}^{\ominus} - D \big) \, M^{-1} \mbox{ .}
\end{equation} 
\end{subequations}
Note that Eq. \eqref{eq:aster} links the solution vector $\tilde{u}_h$ and its time derivative to the two interface variables $\widehat{u}_h$ at the boundaries of the considered element.

The hybridized DG discretization of the linear convection-diffusion equation \eqref{e:lcd} in matrix notation is given by Equations \eqref{eq:III}, \eqref{eq:II} and \eqref{eq:I}; which are required to hold in all elements and all faces. Equations \eqref{eq:star} and \eqref{eq:aster} are an equivalent formulation in terms of $\tilde{u}_h$ and $\widehat{u}_h$ only. Both formulations need to be further equipped with the initial condition $\tilde{u}_h(t=0) = \tilde{u}_{h,0}$, the discretized version of Eq. \eqref{e:lcd2_3}, where the right-hand side is the vector of coefficients of the Galerkin projection of $u_0$ and is given by
\begin{equation}
\tilde{u}_{h,0} = M^{-1} d \mbox{ ,}
\end{equation}
where we have introduced the vector
\begin{equation}
d_j = \int_{\Omega_{\textnormal{ref}}} u_0 \, \phi_j \mbox{ .}
\end{equation}
We note that all the methods within the hybridized DG family, including Hybridizable DG, Embedded DG and Interior Embedded DG, reduce to the same scheme in one-dimensional problems, and therefore no difference between them has been made here.

\subsection{Non-modal analysis formulation}

As in previous works \cite{Mengaldo:ComputersFluids:2017,Mengaldo:eigenFR:2018a,Moura:15a,Moura:16b}, we focus on the analysis of diffusion, which is more relevant and usually dominates dispersion errors in under-resolved turbulence simulations \cite{Ainsworth:2004,Hu:1999,Moura:15a}. In particular, we are concerned about the short-term diffusion properties, in wavenumber space, of the hybridized DG discretization of Eq. \eqref{e:lcd}. That is, if the initial condition is a single Fourier mode $u_0 \propto \exp(i \kappa x)$, where $\kappa \in \mathbb{R}$ denotes the wavenumber, how does the magnitude of the numerical solution evolve over time, and in particular right after $t = 0$? To this end, we define the short-term diffusion as
\begin{equation}
\label{e:lambdaDef}
\varpi^* := \frac{d \log \norm{u_h}}{d \tau^*} \bigg |_{\tau^*=0} \mbox{ ,}
\end{equation}
where $\norm{\, \cdot \, }$ denotes the $L^2$ norm and $\tau^* = \tau \, (P+1) = t \, a \, (P+1) / h$ is a non-dimensional time based on the convection time between degrees of freedom. Note that we define the distance between degrees of freedom as $h^* = h / (P+1)$ and that $\tau^* = 1$ is the time it takes for a flow with speed $a$ to travel a single DOF. The $*$ superscript, such as in $\varpi^*$, $\tau^*$ and $h^*$, is used to indicate that a $(P+1)$ factor has been applied to account for the $P+1$ degrees of freedom per element. Also, we note that Eq. \eqref{e:lambdaDef} can be rewritten as
\begin{equation}
\label{e:lambdaDef_v2}
\varpi^* = \lim_{\tau^* \downarrow 0} \, \frac{1}{\tau^*} \log \Bigg( \frac{\norm{u_h}}{\norm{u_{h,0}}} \Bigg) \mbox{ ,}
\end{equation}
which some readers may find easier to interpret. As we shall see, $\varpi^*$ depends on the wavenumber $\kappa$, the modified P\'eclet number $\textnormal{Pe}^* = |a| \, h^* / \nu = \textnormal{Pe} / (P+1)$ and the details of the hybridized DG scheme, such as the polynomial order $P$ and the upwinding parameter $\beta$.

Intuitively, $\varpi^*$ informs of the decay rate of the numerical solution, per \textit{unit convection time between degrees of freedom}, at early times, starting from the initial condition $\exp(i \kappa x)$. In particular,
\begin{equation}
\norm{u_h(\tau^*)} \approx \norm{u_{h,0}} \exp(\varpi^* \tau^*) 
\end{equation}
at early times, and thus $\exp(\varpi^*)$ can be considered as a \textit{damping factor per DOF crossed}.

Next, we derive an explicit expression for $\varpi^*$. It can be shown (see footnote \footnote{The Galerkin projection of $\exp(i \kappa x)$ trivially features this wave-like behavior, and thus \eqref{e:waveLike_u} holds at $t=0$. Equation \eqref{e:waveLike_uHat} at $t = 0$ then follows from \eqref{eq:star}. Since Equations \eqref{e:waveLike} are satisfied at $t = 0$, it follows from \eqref{e:discrete_cd_Fourier_1} (which holds at any given time under the previous assumptions) that they are also satisfied at all subsequent times $t \geq 0$.}) that if $u_0 \propto \exp(i \kappa x)$, then the relations
\begin{subequations}
\label{e:waveLike}
\begin{equation}
\label{e:waveLike_u}
\tilde{u}_h^L = \tilde{u}_h \exp(-i \kappa h) \mbox{ ,} \qquad \qquad \tilde{u}_h^R = \tilde{u}_h \exp(+i \kappa h) \mbox{ ,} 
\end{equation}
\begin{equation}
\label{e:waveLike_uHat}
\widehat{u}_{h,\ominus} = \widehat{u}_{h,\oplus} \exp(-i \kappa h) \mbox{ ,} 
\end{equation}
\end{subequations}
hold for all elements and all times. Similarly to the notation adopted before for the elements neighboring an interface, we use the superscripts $L$ and $R$ in \eqref{e:waveLike_u} to denote the left and right neighboring elements of a given element. The wave-like behavior of the numerical solution allows to reduce the dimensionality of the problem from countably many (infinite) degrees of freedom to $P+1$ degrees of freedom, and this in turn makes our non-modal analysis possible. In particular, it now follows from Equations \eqref{eq:star} and \eqref{e:waveLike_uHat} that
\begin{equation} \label{eq:starb}
\widehat{u}_h = \left( \tilde{\phi}^T_\oplus B_\oplus \tilde{u}_h^L + \tilde{\phi}^T_\ominus B_\ominus \tilde{u}_h^R \right) b^{-1} \mbox{ ,}
\end{equation}
where $b = b (\kappa h; \mbox{Pe}, P, \bar{\sigma})$ is a scalar defined as
\begin{equation} \label{eq:scalarb}
b = \bar{\sigma} + \left( m_{\ominus}^{\ominus} - m_{\oplus}^{\ominus} \exp(- i \kappa h) - m_{\ominus}^{\oplus} \exp(+i \kappa h) + m_{\oplus}^{\oplus} \right) \, \textnormal{Pe}^{-1} \mbox{ .}
\end{equation}
Inserting \eqref{e:waveLike_u} and \eqref{eq:starb} into \eqref{eq:aster}, one finally obtains
\begin{equation}\label{e:discrete_cd_Fourier_1}
\frac{h}{a} \, \frac{d \tilde{u}_h}{dt} = \tilde{Z}_h \tilde{u}_h , 
\end{equation}
where $\tilde{Z}_h = \tilde{Z}_h (\kappa h; \mbox{Pe}^*, P, \bar{\sigma})$ is a square matrix given by
\begin{equation}
\label{e:Zh_tilde}
\tilde{Z}_h = 2b^{-1} M^{-1} \big( A_{\ominus} \Phi_{\ominus}^{\ominus} B_{\ominus} + A_{\ominus} \Phi_{\oplus}^{\ominus} B_{\oplus} \exp(- i \kappa h) + A_{\oplus} \Phi_{\ominus}^{\oplus} B_{\ominus} \exp(+i \kappa h) + A_{\oplus} \Phi_{\oplus}^{\oplus} B_{\oplus} - A b \big) \mbox{ ,}
\end{equation}
with $\Phi_\ominus^\ominus$ and $\Phi_\oplus^\oplus$ are given by \eqref{eq:matsPhi}, and
\begin{equation} \label{eq:matsPhiNew}
\Phi_{\oplus}^{\ominus} = \tilde{\phi}_{\ominus} \tilde{\phi}_{\oplus}^T \mbox{ ,} \qquad \qquad \Phi_{\ominus}^{\oplus} = \tilde{\phi}_{\oplus} \tilde{\phi}_{\ominus}^T \mbox{ .}
\end{equation}

Since $\varpi^*$ is independent of the choice of basis, we assume without loss of generality that $\phi_j$ is the orthonormal Legendre polynomial of degree $j$ in $\Omega_{\textnormal{ref}} = [-1,1]$; in which case we can obtain a closed-form expression for $\varpi^*$. Combining Equations \eqref{e:lambdaDef} and \eqref{e:discrete_cd_Fourier_1}, using inner product properties, orthonormality of Legendre polynomials and the wave-like behavior of the numerical solution, it follows that
\begin{equation}
\label{e:lambda_expr}
\begin{split}
\varpi^* = & \, \frac{d \log \norm{u_h}}{d \tau^*} \bigg |_{\tau^*=0} = \frac{1}{\norm{u_h}} \, \frac{d \norm{u_h}}{d \tau^*} \bigg |_{\tau^*=0} = \frac{h^*}{a} \, \frac{1}{(\tilde{u}_{h}^{\dagger} \tilde{u}_{h})^{1/2}} \, \frac{d \, (\tilde{u}_{h}^{\dagger}  \tilde{u}_{h})^{1/2}}{d t} \bigg |_{t=0} \\
= & \, \frac{h^*}{2 a} \, \frac{1}{\tilde{u}_{h}^{\dagger} \tilde{u}_{h}} \, \bigg( \frac{d \tilde{u}_h^{\dagger}}{d t} \tilde{u}_h + \tilde{u}_h^{\dagger} \frac{d \tilde{u}_h}{d t} \bigg) \bigg |_{t=0} = \frac{1}{2} \, \frac{1}{P+1} \, \frac{u_{h,0}^{\dagger} \, \tilde{Z}_h^{\dagger} \, u_{h,0} + u_{h,0}^{\dagger} \, \tilde{Z}_h \, u_{h,0}}{\tilde{u}_{h,0}^{\dagger} \, \tilde{u}_{h,0}} \\
= & \, \frac{1}{P+1} \, \mathbb{R}e \Bigg[ \frac{\tilde{u}_{h,0}^{\dagger} \, \tilde{Z}_h \, \tilde{u}_{h,0}}{\tilde{u}_{h,0}^{\dagger} \, \tilde{u}_{h,0}} \Bigg] \mbox{ .} 
\end{split}
\end{equation}
where the $\dagger$ superscript denotes conjugate transpose and $\mathbb{R}e$ the real part of a complex number. Note that the value of $\varpi^*$ is independent of the amplitude of the Fourier mode. Taking $u_0 = \exp(i \kappa x)$, it follows that $\tilde{u}_{h,0} = \alpha = \alpha(\kappa h)$ with
\begin{subequations}
\label{e:alpha}
\begin{alignat}{2}
\label{e:alpha_1}
\alpha_0 = \sqrt{2} \, \frac{\sin z}{z} \mbox{ ,} \qquad \qquad \alpha_1 = \frac{i \sqrt{6}}{z} \bigg( \frac{\sin z}{z} - \cos z \bigg) \mbox{ ,} \\
\intertext{and, for $j \geq 1$,}
\label{e:alpha_2}
\alpha_{j+1} = \frac{\sqrt{4 j+6}}{z} \, \bigg( m_j \sin z + i \bigg( \frac{\sin z}{z} - \cos z \bigg) m_{j+1} + i \sum_{k=1}^j \sqrt{k+1/2} \, m_{j+k+1} \, \alpha_k \bigg) \mbox{ ,}
\end{alignat}
\end{subequations}
where $z = \kappa h / 2$ and $m_j = \textnormal{mod}(j,2)$ is the modulus of $j$ after division by two. Equations \eqref{e:alpha} hold only for orthonormal Legendre polynomials \cite{Moura:15a}. Equations \eqref{e:Zh_tilde}, \eqref{e:lambda_expr} and \eqref{e:alpha} provide a closed-form expression for $\varpi^* = \varpi^* (\kappa h; \mbox{Pe}^*, P, \bar{\sigma})$.

We note that $\tilde{Z}_h$ is the same matrix (up to complex sign) as that in eigenanalysis \cite{Moura:18a}. The difference between eigenanalysis and our non-modal analysis is that the former concerns about the modal behavior (i.e.\ the eigenvalues and eigenmodes) of $\tilde{Z}_h$, whereas we concern about its non-modal behavior (i.e.\ we consider the contribution of all the eigenmodes for each wavenumber), and in particular about its non-modal short-term dynamics. Also, our definition of short-term diffusion $\varpi^*$ is consistent with the definition of diffusion introduced in eigenanalysis \cite{Moura:15a}: If the Fourier mode is an eigenmode of $\tilde{Z}_h$, as it is the case for $P=0$, then $(P+1)\, \varpi^*$ coincides with the real part of the corresponding eigenvalue, and non-modal analysis reconciles with eigenanalysis. Similarly, if our evaluation is applied to the primary eigenmodes (instead of to the Fourier modes as described above), our results will match the primary eigencurves in \cite{Moura:15a}.

{\bf Remark:} Strictly speaking, Equations \eqref{e:lambdaDef} and \eqref{e:lambdaDef_v2} should read as
\begin{equation}
\label{e:beingPicky}
\varpi^* := \lim_{n \to \infty} \frac{d \log \norm{\chi_{[-n,n]} \, u_h}}{d \tau^*} \Bigg |_{\tau^*=0} \mbox{ ,} \qquad \qquad \varpi^* = \lim_{n \to \infty} \lim_{\tau^* \downarrow 0} \frac{1}{\tau^*} \log \Bigg( \frac{\norm{\chi_{[-n,n]} \, u_h}}{\norm{\chi_{[-n,n]} \, u_{h,0}}} \Bigg) \mbox{ ,}
\end{equation}
respectively, where $\chi_{[-n,n]}$ denotes the indicator function in $[-n,n]$, in order for the norms and thus $\varpi^*$ to be well-defined. 
A limiting process is also required, and intentional abuse of notation is used, whenever $\norm{u_h}$ is written outside of a logarithm.

\subsection{Non-modal analysis results}

We present the non-modal analysis results through the so-called short-term diffusion curves. For given $\textnormal{Pe}^*$, $P$ and $\beta$, the short-term diffusion curves show $\varpi^*$ ($y$ axis) as a function of the non-dimensional wavenumber $\kappa h^* = \kappa h / (P+1)$ ($x$ axis). The right limit of the $x$ axis corresponds to the Nyquist wavenumber of the grid, defined as $\kappa_N = \pi / h^*$, and thus $\kappa_N h / (P+1) = \pi$. The exact diffusion curves are indicated with dashed lines.

Before presenting the short-term diffusion curves, we briefly discuss on how these curves \textit{should look like} from the perspectives of robustness and accuracy. For robustness purposes, monotonic ($d \varpi^* / d \kappa \leq 0$) and slowly-varying curves are preferred as they lead to more stable discretizations, particularly for nonlinear systems due to the nonlinear interactions between wavenumbers. Regarding accuracy, the short-term diffusion should agree as much as possible with the exact diffusion curve in the case of well-resolved simulations, including direct numerical simulation (DNS) of turbulent flows. However, for under-resolved computations of systems featuring a kinetic energy cascade, such as in LES, additional diffusion is desired at large wavenumbers in order for the eddy-viscosity effect of the missing scales to be somehow accounted for by the numerics. Also for robustness purposes, additional numerical diffusion at large wavenumbers is beneficial so as to provide further regularization (i.e.\ damp oscillations) and avoid energy accumulation at the smallest scales captured. These considerations are based on \textit{a priori} knowledge and \textit{a posteriori} insights from the numerical results in Section \ref{s:applicationNonLinear}.

\subsubsection{Effect of the polynomial order}

\begin{figure}[t]
\centering
\includegraphics[width=0.45\textwidth]{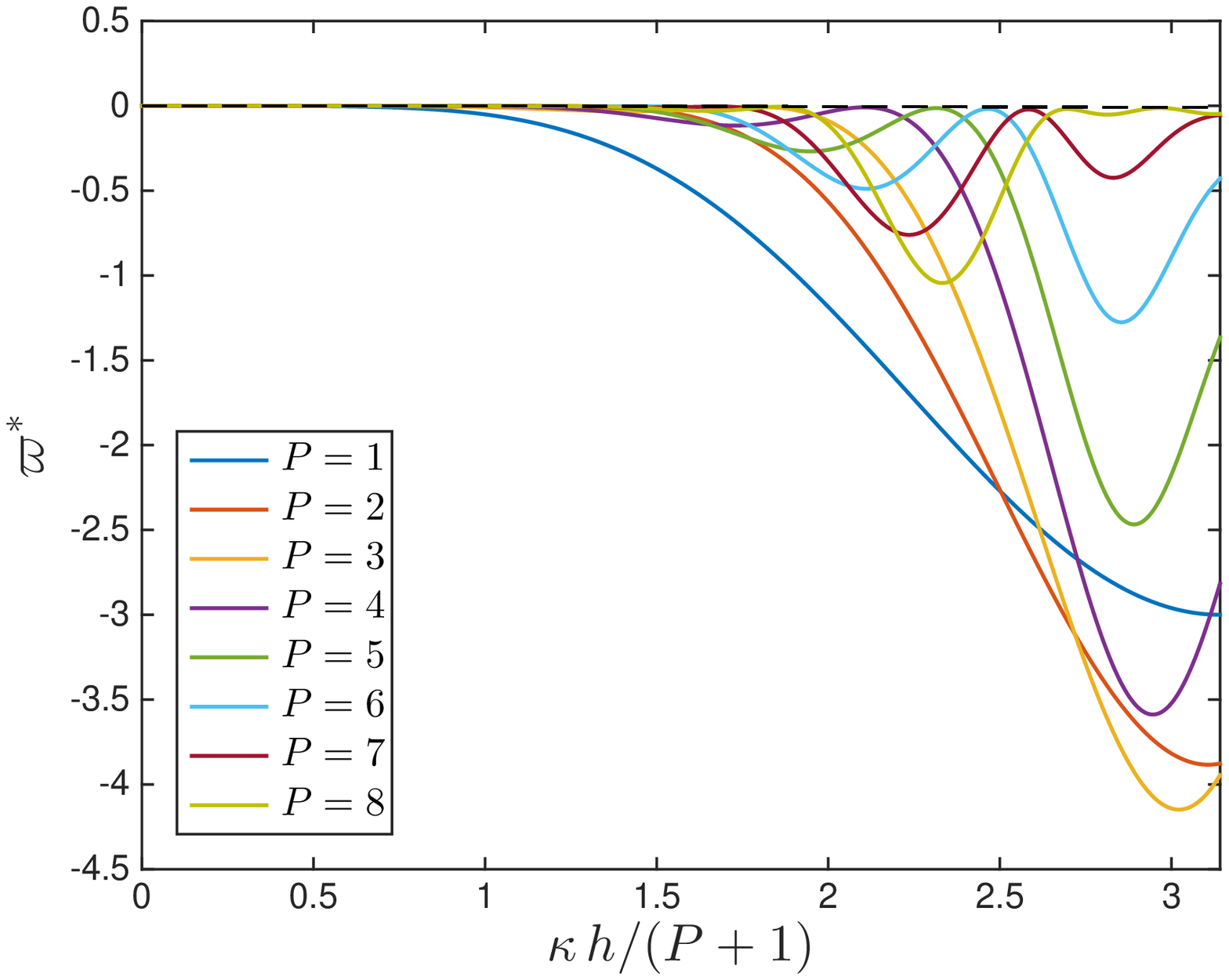}
\hfill \includegraphics[width=0.45\textwidth]{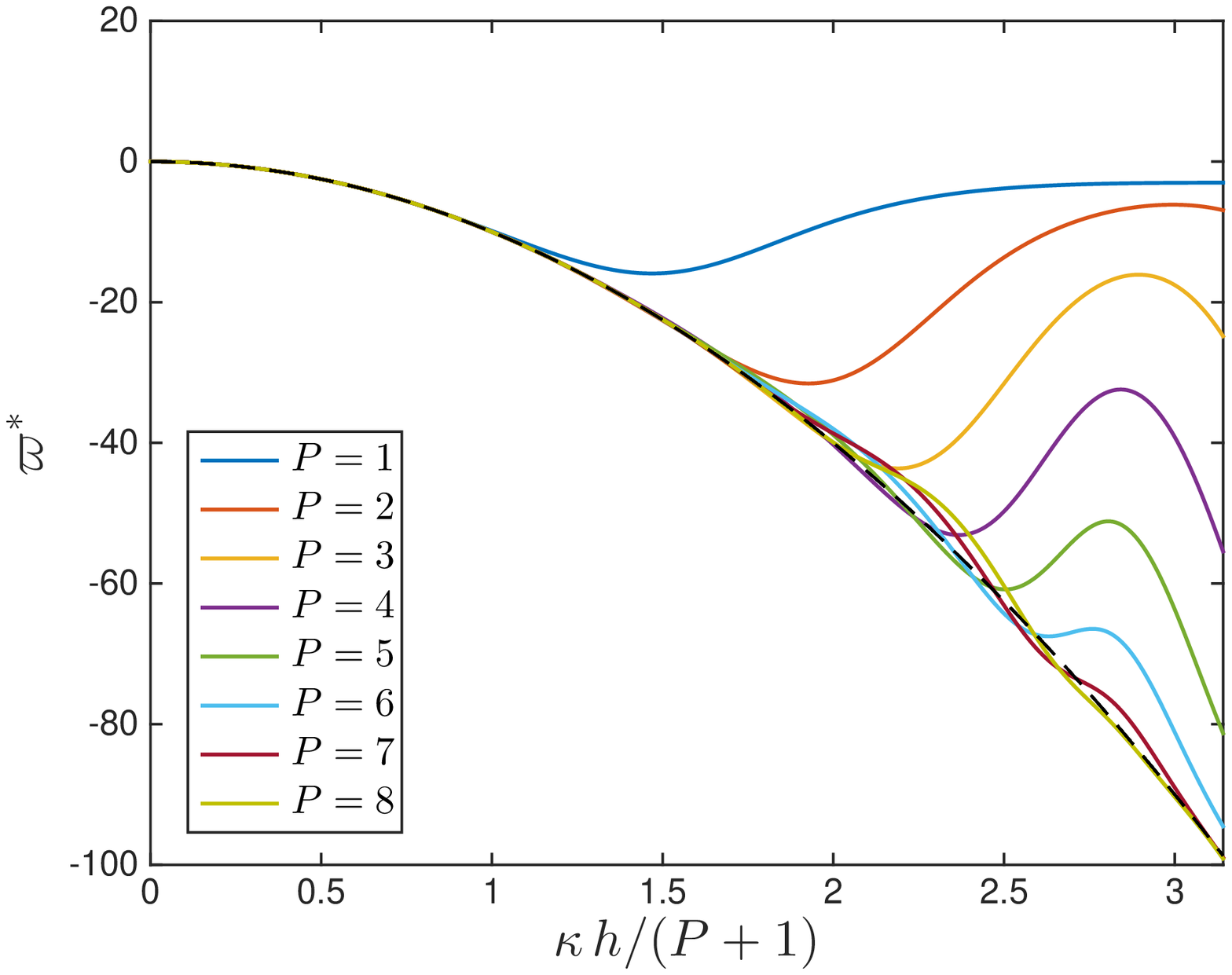}
\caption{\label{f:shortTimeDiffusionCurves_pStudy} Short-term diffusion curves as a function of the polynomial order for $\beta = 1$ in convection-dominated $\textnormal{Pe}^* = 10^3$ (left) and diffusion-dominated $\textnormal{Pe}^* = 0.1$ (right) regimes. The exact solution of the convection-diffusion equation is shown in dashed black line. Note a different scale in the $y$ axis is used for each figure.}
\end{figure}

Figure \ref{f:shortTimeDiffusionCurves_pStudy} shows the short-term diffusion curves as a function of the polynomial order for standard upwinding $\beta = 1$ in convection-dominated $\textnormal{Pe}^* = 10^3$ (left) and diffusion-dominated $\textnormal{Pe}^* = 0.1$ (right) regimes. 
The exact solution of the convection-diffusion equation is shown in dashed black line.

For convection-dominated flows, high polynomial orders lead to non-monotonic short-term diffusion characteristics. In particular, very small diffusion is introduced at some specific wavenumbers. As discussed before and shown by the numerical results in Sections \ref{s:burgers} and \ref{s:TGV}, this may lead to nonlinear instabilities. Also, we recall that some amount of numerical diffusion near $\kappa_N$ (preferably monotonic in wavenumber space) is desired, both for accuracy and robustness, in under-resolved turbulence simulations in order to replicate the dissipation that takes place in the subgrid scales. The short-term diffusion properties that are better suited, both in terms of accuracy per DOF and robustness, for convection-dominated under-resolved turbulence simulations, seem to be those for polynomial orders $P=2$, $3$ and $4$. For $P=1$, diffusion is introduced at scales  that are much larger than the Nyquist wavenumber. We note that convection-dominated, from the cell P\'eclet number perspective, is the regime most commonly encountered in large-eddy simulation. For diffusion-dominated problems, higher $P$ improves both accuracy per degree of freedom and robustness.

\subsubsection{Effect of the P\'eclet number}

\begin{figure}[t]
\centering
\includegraphics[width=0.45\textwidth]{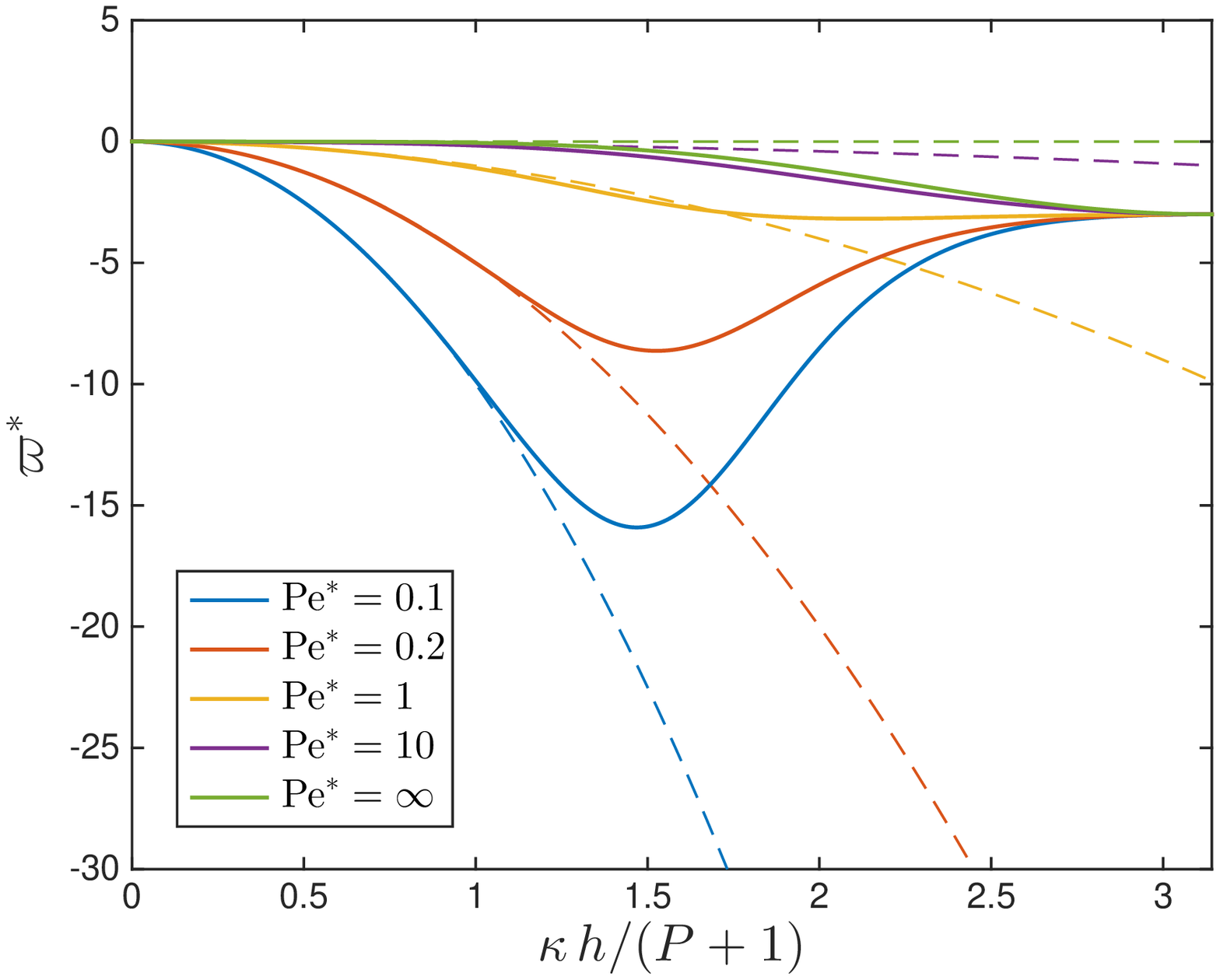}
\hfill \includegraphics[width=0.45\textwidth]{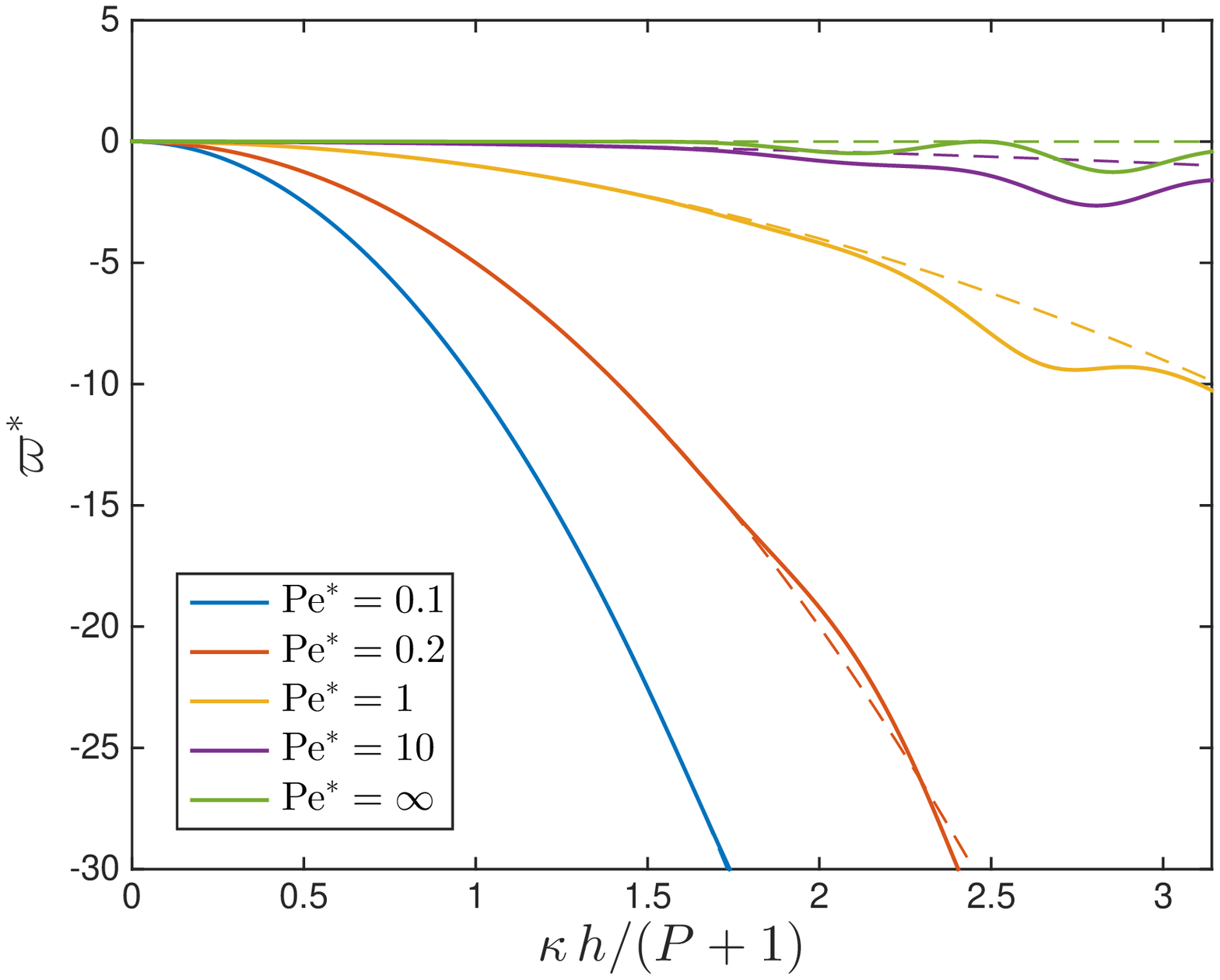}
\caption{\label{f:shortTimeDiffusionCurves_iPeStudy} Short-term diffusion curves as a function of the P\'eclet number for standard upwinding $\beta = 1$ and polynomial orders $P=1$ (left) and $P=6$ (right). The exact solution of the convection-diffusion equation is shown in dashed lines.}
\end{figure}

Figure \ref{f:shortTimeDiffusionCurves_iPeStudy} shows the short-term diffusion curves as a function of the P\'eclet number for standard upwinding $\beta = 1$ with polynomial orders $P=1$ (left) and $P=6$ (right). The exact solution of the convection-diffusion equation is shown in dashed lines. As noted in the polynomial order study, high $P$ schemes are better suited to diffusion-dominated problems, both in terms of accuracy and robustness. As for low polynomial orders, moderately high wavenumbers are poorly resolved regardless of the P\'eclet number. Robustness of low $P$ schemes seems to improve in the convection-dominated regime.

\begin{figure}[t]
\centering
\includegraphics[width=0.45\textwidth]{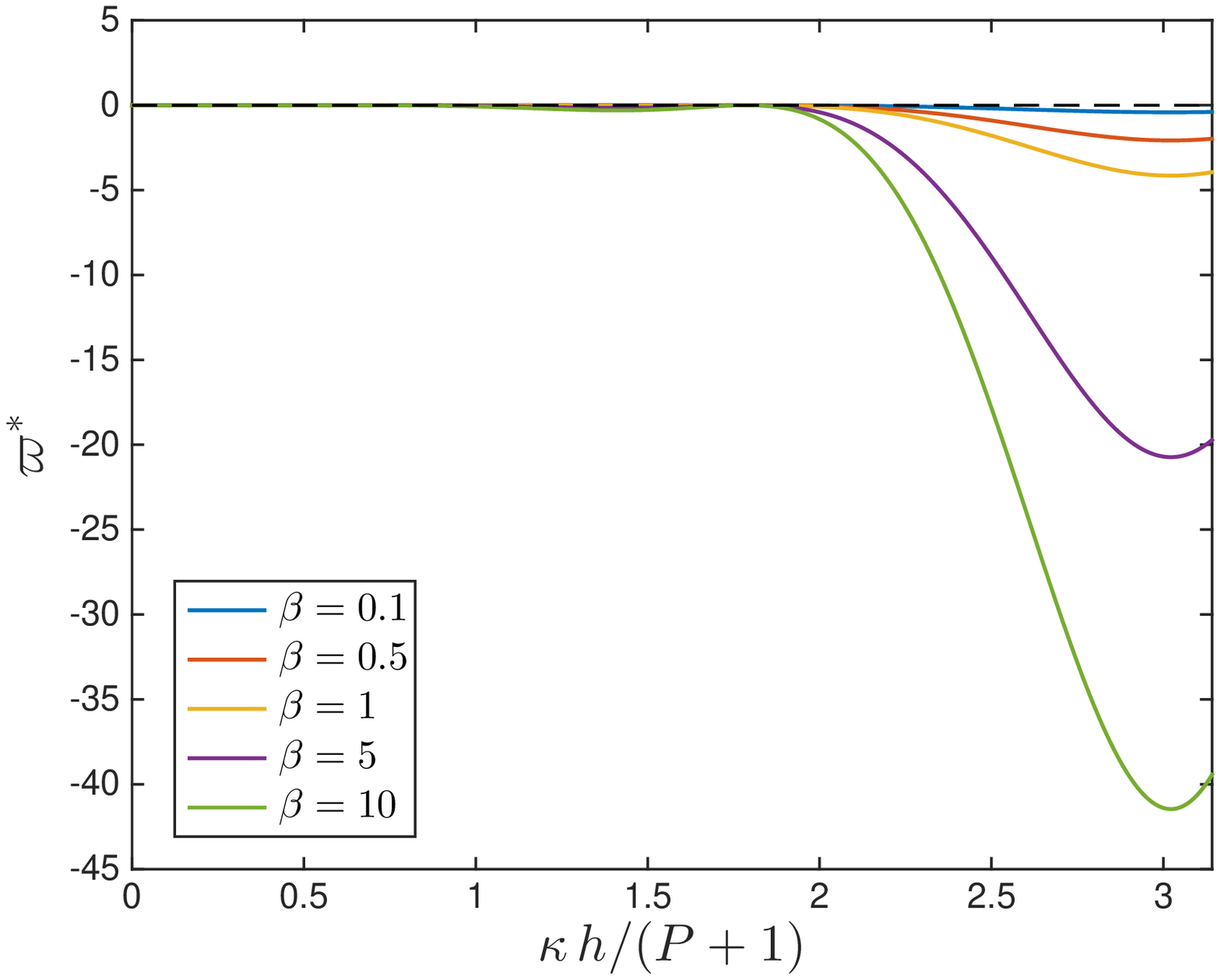}
\hfill \includegraphics[width=0.45\textwidth]{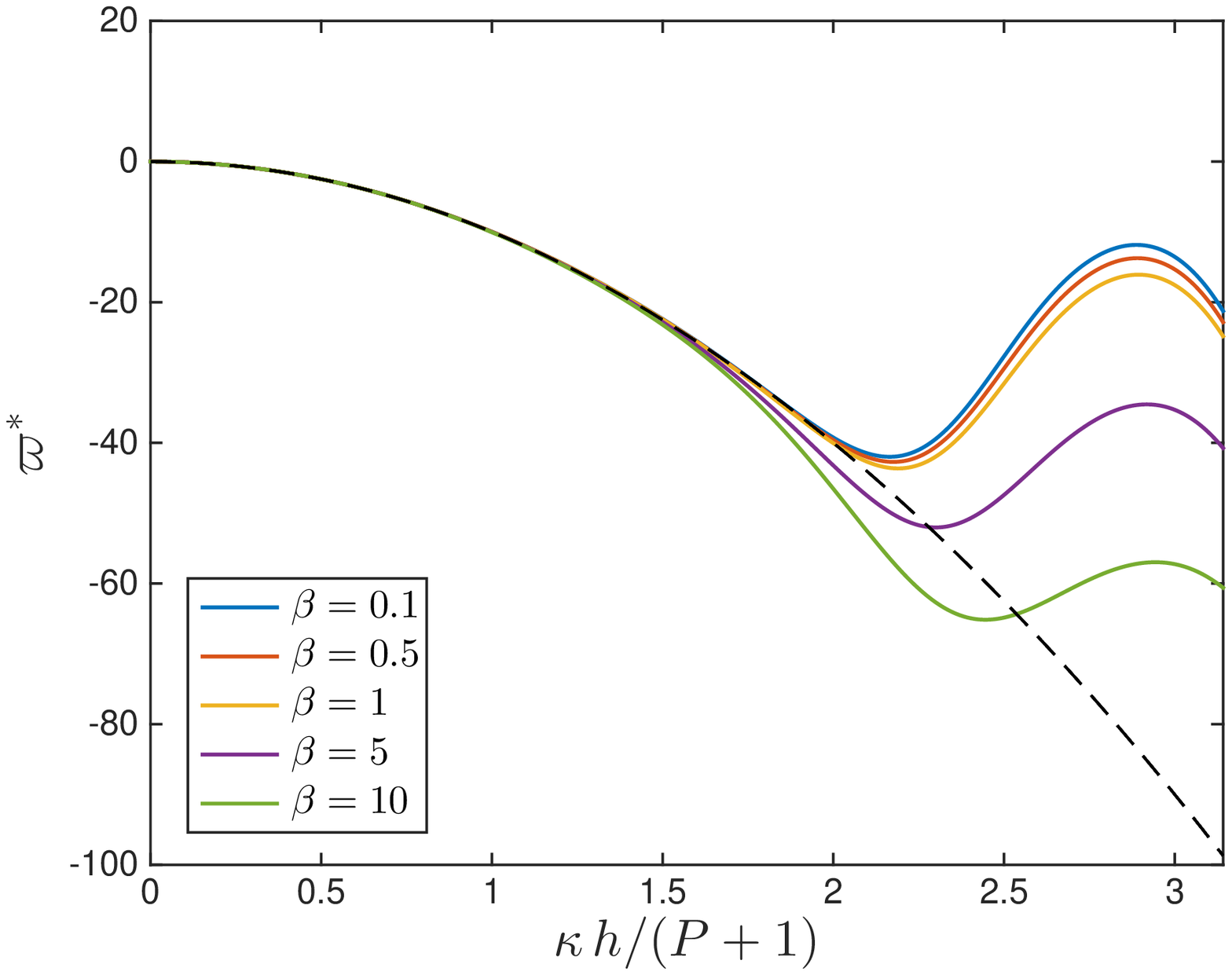}
\caption{\label{f:shortTimeDiffusionCurves_betaStudy} Short-term diffusion curves as a function of the upwinding parameter for $P = 3$ in convection-dominated $\textnormal{Pe}^* = 10^3$ (left) and diffusion-dominated $\textnormal{Pe}^* = 0.1$ (right) regimes. The exact solution of the convection-diffusion equation is shown in dashed black line. Note a different scale in the $y$ axis is used for each figure.}
\end{figure}

\subsubsection{Effect of the upwinding parameter}

Figure \ref{f:shortTimeDiffusionCurves_betaStudy} shows the short-term diffusion curves as a function of the upwinding parameter for $P=3$ in convection-dominated $\textnormal{Pe}^* = 10^3$ (left) and diffusion-dominated $\textnormal{Pe}^* = 0.1$ (right) regimes. The exact solution of the convection-diffusion equation is shown in dashed black line. We recall the upwinding parameter was introduced and defined in Eq. \eqref{eq:flux_}.

For convection-dominated regimes, strong under/over-upwinding is to be avoided as it causes dissipation at large wavenumbers to rise too slow/fast. The former leads to a lack of small-scale regularization, and the latter causes the bottleneck phenomenon and its associated energy bump, detrimental to both solution quality and numerical stability (see \cite{moura2017eddy}). Nevertheless, it may be the case that a controlled level of under/over-upwinding may be useful for certain simulations, e.g.\ when the eddy-viscosity effect of the missing scales is not represented correctly by the standard upwind condition. For diffusion-dominated problems, the scheme benefits form over-upwinding; which is not completely surprising since no explicit stabilization has been used for the diffusion operator.

\section{\label{s:applicationNonLinear}Nonlinear problems}

In order to assess how the non-modal analysis results extend to the nonlinear setting, we apply it to the Burgers, Euler and Navier-Stokes equations. One-dimensional and three-dimensional examples are considered.

\subsection{\label{s:burgers}Application to the Burgers equation}

\subsubsection{Problem description}

We consider the following one-dimensional forced Burgers turbulence problem \cite{Chekhlov:1995}
\begin{subequations}
\label{e:burgersTurb}
\begin{alignat}{2}
\label{e:burgersTurb1}
& \frac{\partial u}{\partial t} + \frac{1}{2} \frac{\partial u^2}{\partial x} = \frac{A_F}{\sqrt{\Delta t}} \sum_{N \in \mathbb{N}_F} \frac{\sigma_N(t)}{\sqrt{|N|}} \exp \bigg( i \frac{2 \pi N}{L} x \bigg) , \qquad & t > 0 , \\
& u|_{x=-L/2} = u|_{x=L/2} , \qquad & t \geq 0 , \\
& u = u_0 , \qquad & t = 0 , 
\end{alignat}
\end{subequations}
where $u_0 > 0$ denotes the initial velocity (constant in the domain), $L$ is the length of the computational domain $\Omega$, $\Delta t$ is the time-step size used in the simulation, $A_F$ is an amplitude constant for the forcing term, $\mathbb{N}_F = \{ \pm 1 , \dots , \pm N_c \}$ is a collection of integers, and $\sigma_{N}$ is a standard Gaussian random variable (zero mean and unit variance) that is independent for each wavenumber and each time step. We set $N_c=80$ and $A_F = \sqrt{8} \cdot 10^{-1} \, u_0^{3/2} \, L^{-1/2}$ for the numerical experiments in this section. This completes the non-dimensional description of the problem.

The choice of forcing in \eqref{e:burgersTurb} yields, for wavenumbers below the cut-off wavenumber $\kappa_c = 2 \pi N_c / L$, a $-5/3$ slope for the inertial range of the energy spectrum \cite{Adams:2004,Chekhlov:1995,Moura:15a,Zikanov:1997} and thus resembles Navier-Stokes turbulence within the Burgers setting. As customary in the literature, we use the term Burgers \textit{turbulence} to refer to the chaotic and turbulent-like behavior featured by the solution of the Burgers equation.

\subsubsection{Details of the numerical discretization}

We use the hybridized DG method with various polynomial orders to discretize Eq. \eqref{e:burgersTurb} in space. We recall that HDG, EDG, IEDG and all other schemes within the hybridized DG family reduce to the same scheme in one-dimensional problems, and there is only one type of hybridized DG method for this problem. We refer the interested reader to \cite{Nguyen:09:nonlinear} for the details of the hybridized DG discretization of the one-dimensional Burgers equation. The upwinding parameter is $\beta = 1$ (and thus $\sigma = |\widehat{u}_h|$), and the total number of degrees of freedom is $N_{DOF} = (P+1) \cdot \lceil 1024 / (P+1) \rceil \approx 1024$, where $\lceil \ \  \rceil$ denotes the rounding of a positive real number to the closest larger (or equal) integer. Note this is required in order to obtain an integer number of elements in $\Omega$. Exact integration is used both for the Burgers flux and the forcing term. For the former, Gauss-Legendre quadrature with the required number of points to ensure exact integration of polynomials of degree $3 \, P$ (and thus of the Burgers flux term in the hybridized DG discretization) is used. The Galerkin projection of the forcing term is integrated exactly using the analytical expressions in \cite{Moura:15a}. The third-order, three-stage $L$-stable diagonally implicit Runge-Kutta DIRK(3,3) scheme \cite{Alexander:77} is used for the temporal discretization with CFL number based on the initial velocity of $\textnormal{CFL} = u_0 \, \Delta t / h^* = 0.05$. The solution is computed from $t = 0$ to $t = 8 \, L / u_0$.

\subsubsection{Numerical results}

\begin{figure}
\centering
\includegraphics[width=1.0\textwidth]{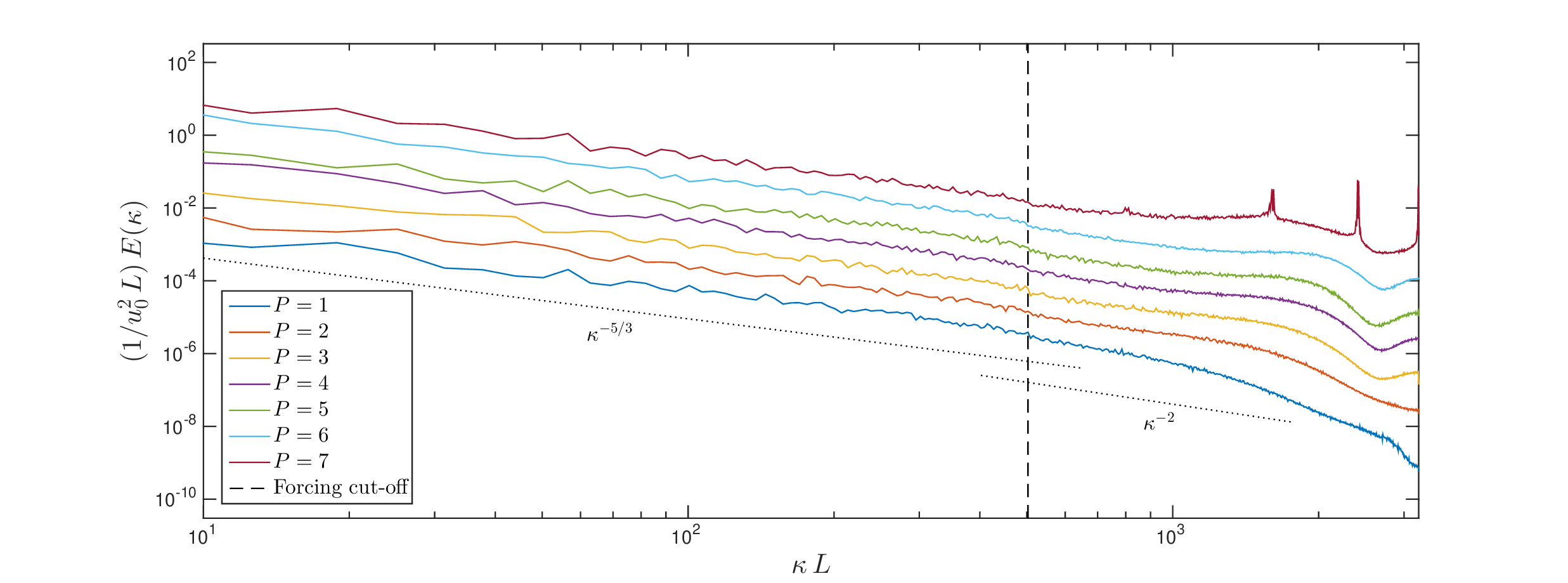}
\hfill \includegraphics[width=1.0\textwidth]{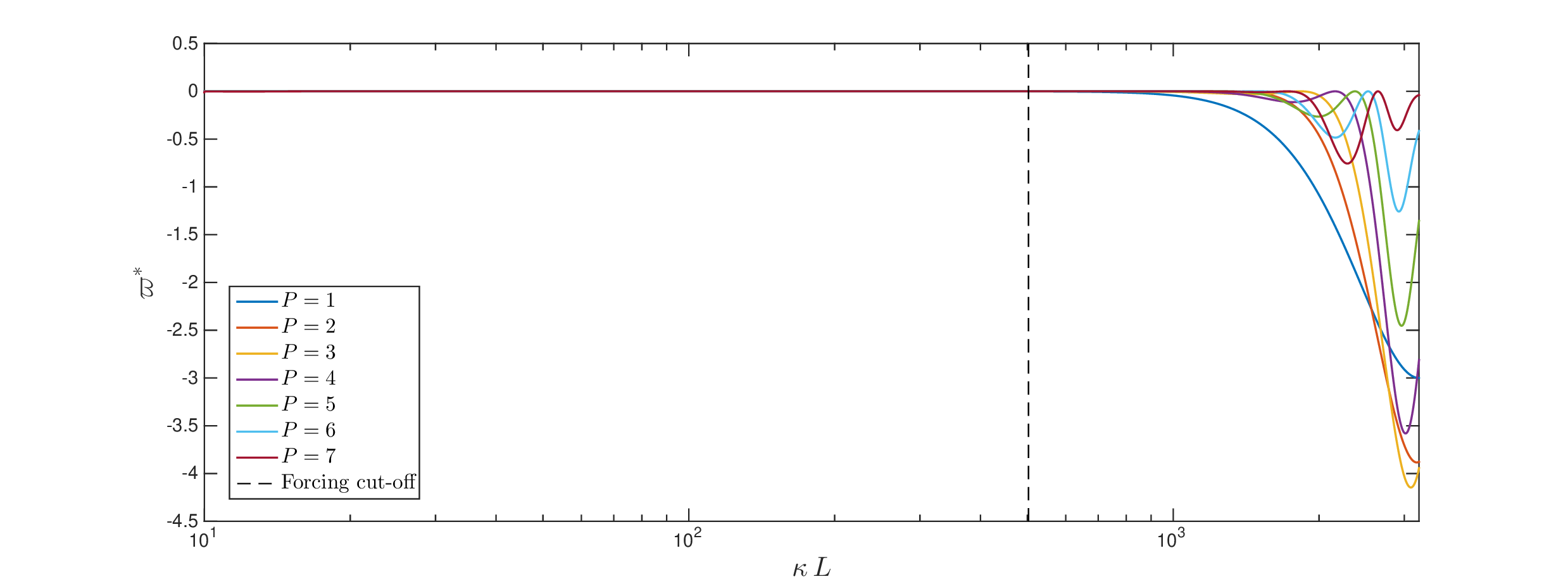}
\caption{\label{f:spectraBurgers} Results for the Burgers turbulence problem. Top: Time-averaged energy spectra from $t = 2 \, L / u_0$ to $t = 8 \, L / u_0$ for $P = 1 , \dots , 7$. The spectra are shifted up by a factor of $4^{P-1}$ to allow for easier visualization. Bottom: Short-term diffusion curves from non-modal analysis. The right limit of the $x$ axis corresponds to the Nyquist wavenumber of the grid.}
\end{figure}

The time-averaged kinetic energy spectra from $t = 2 \, L / u_0$ to $t = 8 \, L / u_0$ for $P = 1 , \dots , 7$ are shown on the top of Figure \ref{f:spectraBurgers}. Polynomial orders $P \geq 8$ failed to converge for this problem. A time refinement study confirmed that the breakdown for the high-order cases occurs independently of the time-step size, and it is therefore attributed to the lack of stability in the hybridized DG discretization; which we note is not provable $L^2$ stable even though exact integration is used. The spectra are shifted up by a factor of $4^{P-1}$ to allow for easier visualization. 
All the spectra feature an inertial range of turbulence with slope $-5/3$ up to $\log_{10} (\kappa_c \, L) = \log_{10} (2 \pi N_c) \approx 2.7$, as expected from the forcing strategy adopted. After the cut-off wavenumber, a slope of $-2$, typical of unforced Burgers turbulence \cite{Bec:2007}, takes place whenever numerical dissipation is still small enough over these wavenumbers. In all the simulations, numerical dissipation eventually becomes significant and affects the shape of the energy spectra near the Nyquist wavenumber of the grid $\kappa_N \, L = \pi N_{DOF} \approx 3217$; which corresponds to the right limit of the $x$ axis.

The short-term diffusion curves from non-modal analysis are shown in the bottom of Figure \ref{f:spectraBurgers}, where the $x$ axis has been mapped from $\kappa \, h / (P+1)$ (as in Figures \ref{f:shortTimeDiffusionCurves_pStudy}$-$\ref{f:shortTimeDiffusionCurves_betaStudy}) to $\kappa \, L$ to facilitate the comparison with the energy spectra. Note that $\textnormal{Pe}^* = \infty$ in this problem due to the lack of physical viscosity. The trends observed in the energy spectra are consistent with non-modal analysis results. First, a numerically induced dissipation range near $\kappa_N$ is observed in the spectrum of the $P = \{ 1 , 2 \}$ and, to a lessen extent, $P=3$ discretizations; which is consistent with the large short-term diffusion of these schemes right before the Nyquist wavenumber. Second, bottlenecks in the turbulence cascade (in the sense of energy accumulations at some specific wavenumbers) are observed for the high $P$ discretizations; which is consistent with the non-monotonicity in the short-term diffusion curves. In particular, the spikes in the spectrum for $P=7$ nearly coincide with those wavenumbers where numerical dissipation, as estimated from non-modal analysis, is approximately zero, and hints as to why instabilities can occur and higher-order cases fail to converge. Note the instabilities for $P \geq 8$ occur even though exact integration is being performed. 
Note also that these results differ from those obtained with standard DG in \cite{Moura:15a} since an `eigenfilter' was applied in that work to the forcing term in order to eliminate the effect of secondary eigenmodes, whereas now all eigenmodes contribute to the dissipation. 
This highlights the relevance of our non-modal analysis for more complex applications, where a precise eigenfiltering is unfeasible.

\subsection{\label{s:TGV}Application to the Euler and Navier-Stokes equations. The Taylor-Green vortex}

\subsubsection{Problem description}

The Taylor-Green vortex (TGV) problem \cite{Taylor:37} describes the evolution of the fluid flow in a three-dimensional cubic domain $\Omega = [-L \pi, L \pi)^3$ with triple periodic boundaries, starting from the smooth initial condition
\begin{equation}
\label{initialCondTGV}
\begin{split}
\rho & = \rho_0 , \\
u_1 & = U_0 \sin \Big( \frac{x}{L} \Big) \cos \Big( \frac{y}{L} \Big) \cos \Big( \frac{z}{L} \Big) , \\
u_2 & = - U_0 \cos \Big( \frac{x}{L} \Big) \sin \Big( \frac{y}{L} \Big) \cos \Big( \frac{z}{L} \Big) , \\
u_3 & = 0 , \\
p & = p_0 + \frac{\rho_0 \, U_0^2}{16} \, \bigg( \cos \Big( \frac{2x}{L} \Big) + \cos \Big( \frac{2y}{L} \Big) \bigg) \, \bigg( \cos \Big( \frac{2z}{L} \Big) + 2 \bigg) , 
\end{split}
\end{equation}
where $\rho$, $p$ and $(u_1, u_2, u_3)$ denote density, pressure and the velocity vector, respectively, and $\rho_0 , \, p_0 , \, U_0 > 0$ are some reference density, pressure and velocity magnitude. Governed by the Navier-Stokes equations (Euler equations in the inviscid case), the large-scale eddy in the initial condition leads to smaller and smaller structures through vortex stretching. For Reynolds numbers $\textnormal{Re} = \rho_0 \, U_0 \, L / \mu$ below about $1000$, where $\mu$ denotes the dynamic viscosity of the fluid, the flow remains laminar at all times \cite{Brachet:91}. Above this threshold, the vortical structures eventually break down and the flow transitions to turbulence\footnote{While no temporal chaos (chaotic attractor) exists in the Taylor-Green vortex, we use the term \textit{turbulence} to refer to the phase of spatial chaos (spatial decoherence) that takes place after $t \approx 7 - 9 \, L / U_0$ for Reynolds numbers above about $1000$.}. After transition, the turbulent motion dissipates all the kinetic energy, and the flow eventually comes to rest through a decay phase similar to that in decaying homogeneous isotropic turbulence, yet not isotropic here. In the high Reynolds number limit (inviscid TGV), there is no decay phase due to the lack of viscosity and the smallest turbulent scales thus become arbitrarily small as time evolves.

In order to investigate different P\'eclet numbers and flow regimes, we consider the Reynolds numbers $100$, $400$, $1600$ and $\infty$. The reference Mach number is set to $\textnormal{Ma} = U_0 / c_0 = 0.1$ in all cases in order to render the flow nearly incompressible, where $c_0$ denotes the speed of sound at temperature $T_0 = p_0 / (\gamma - 1) \, c_v \, \rho_0$. The fluid is assumed to be Newtonian, calorically perfect, in thermodynamic equilibrium, and with Fourier's law of heat conduction and the Stokes' hypothesis. The dynamic viscosity $\mu$ is constant, the Prandtl number $\textnormal{Pr} = 0.71$ and the ratio of specific heats $\gamma = c_p / c_v = 1.4$. This completes the non-dimensional description of the problem.

\subsubsection{Details of the numerical discretization}

The computational domain is partitioned into a uniform $64 \times 64 \times 64$ Cartesian grid and the Embedded DG (EDG) scheme with $P=2$ is used for the spatial discretization. The details of the EDG discretization of the Euler and Navier-Stokes equations are presented in \cite{Peraire:11}. We consider stabilization matrices of the form
\begin{equation}
\bm{\sigma} = \beta \, |\bm{A}_n (\widehat{\bm{u}}_h)| , 
\end{equation}
with $\beta = 0.25$ (under-upwinding) and $\beta = 1.00$ (standard upwinding), and where $\bm{A}_n = \partial (\bm{F} \cdot \bm{n}) / \partial \bm{u}$ denotes the Jacobian matrix of the inviscid flux normal to the element face. We note that the stabilization matrix implicitly defines the Riemann solver in hybridized DG methods, and in particular a Roe-type solver is recovered in the case $\beta = 1.00$. The interested reader is referred to \cite{Fernandez:AIAA:17a,Fernandez:PhD:2018} for additional details on the relationship between the stabilization matrix and the resulting Riemann solver. The third-order, three-stage $L$-stable diagonally implicit Runge-Kutta DIRK(3,3) scheme \cite{Alexander:77} is used for the temporal discretization with CFL number $U_0 \, \Delta t / h^* = 0.1$. The solution is computed from $t = 0$ to $t = 15 \, L / U_0$.

\subsubsection{Numerical results}

\begin{figure}
\centering
\includegraphics[width=0.45\textwidth]{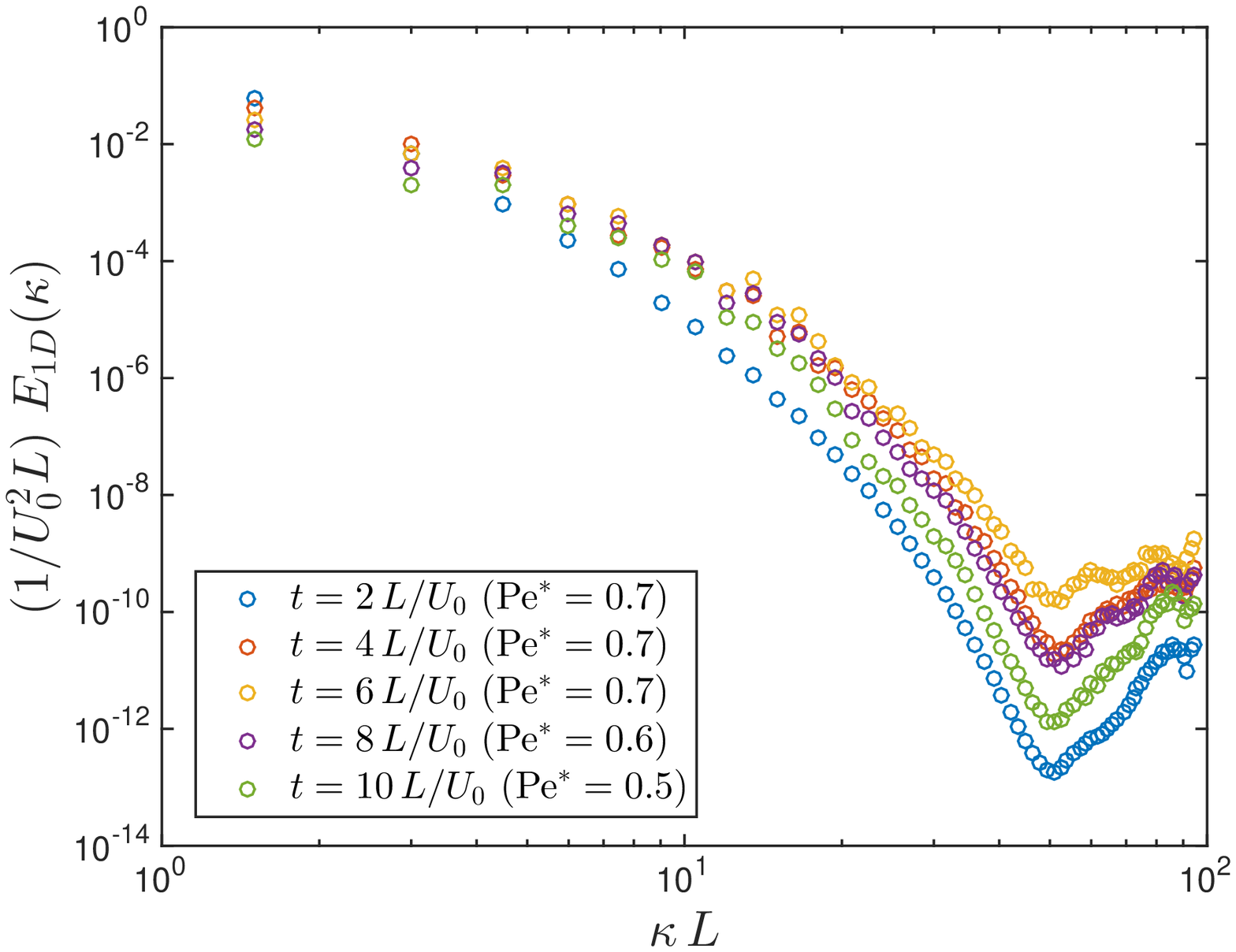}
\hfill \includegraphics[width=0.45\textwidth]{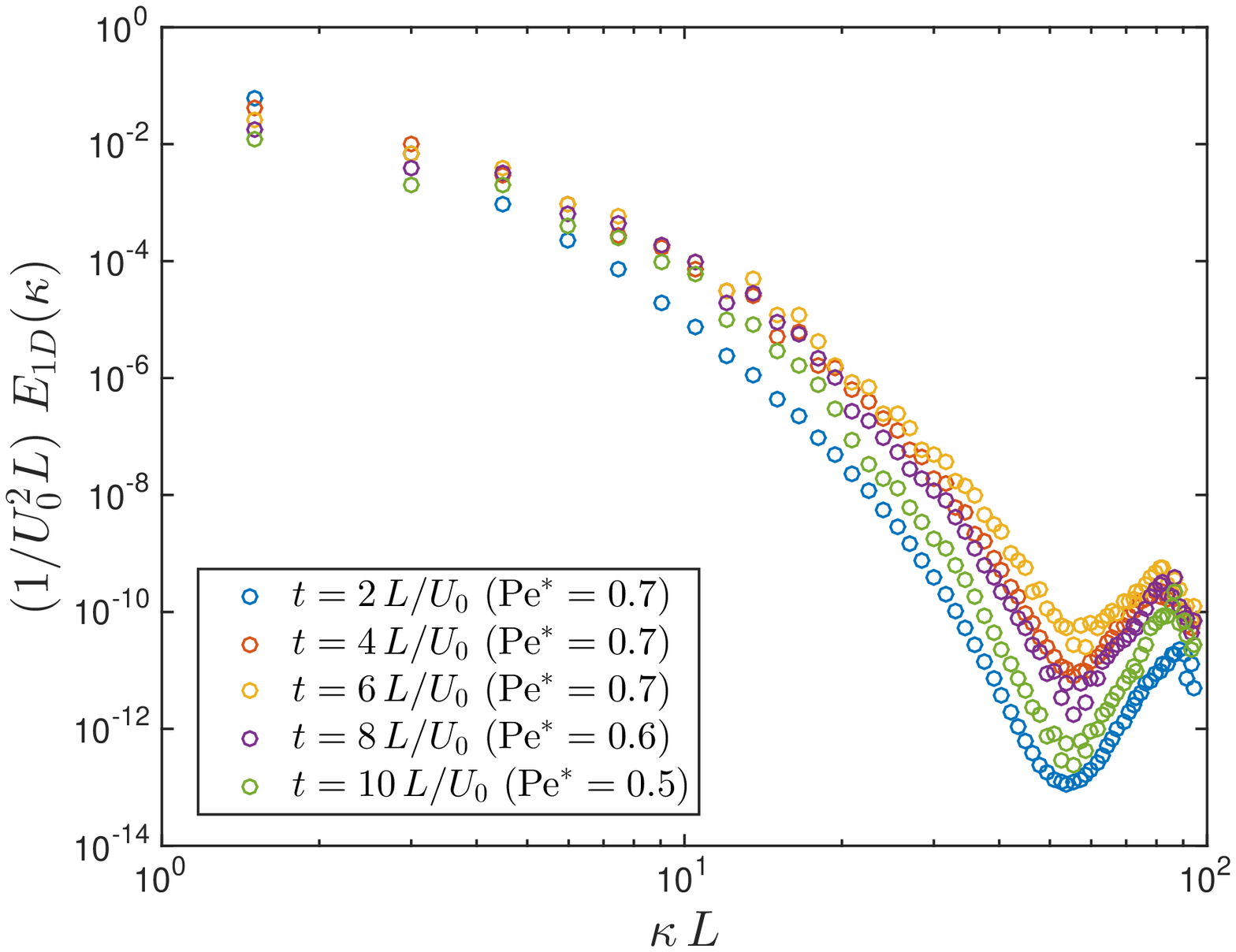}
\hfill \includegraphics[width=0.45\textwidth]{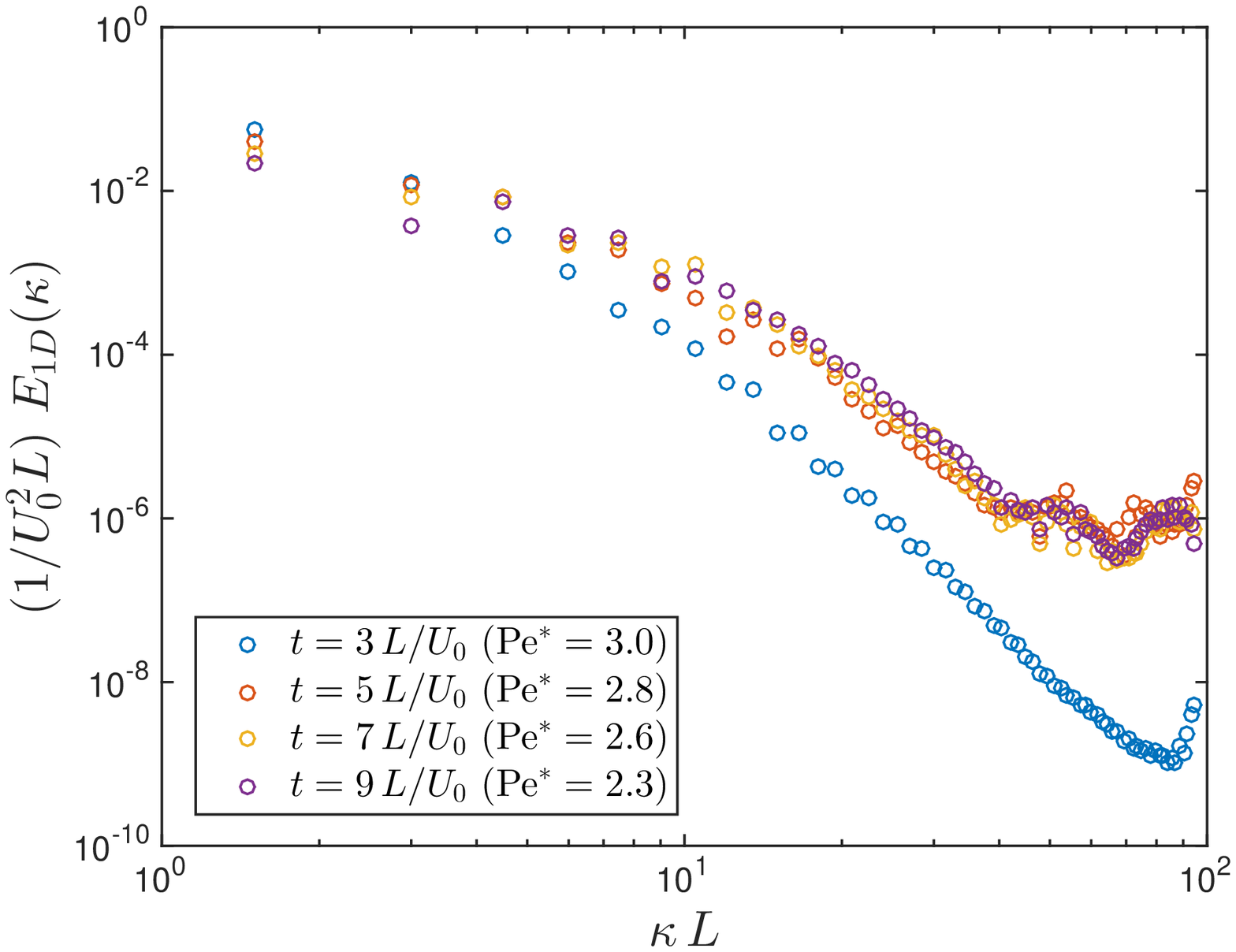}
\hfill \includegraphics[width=0.45\textwidth]{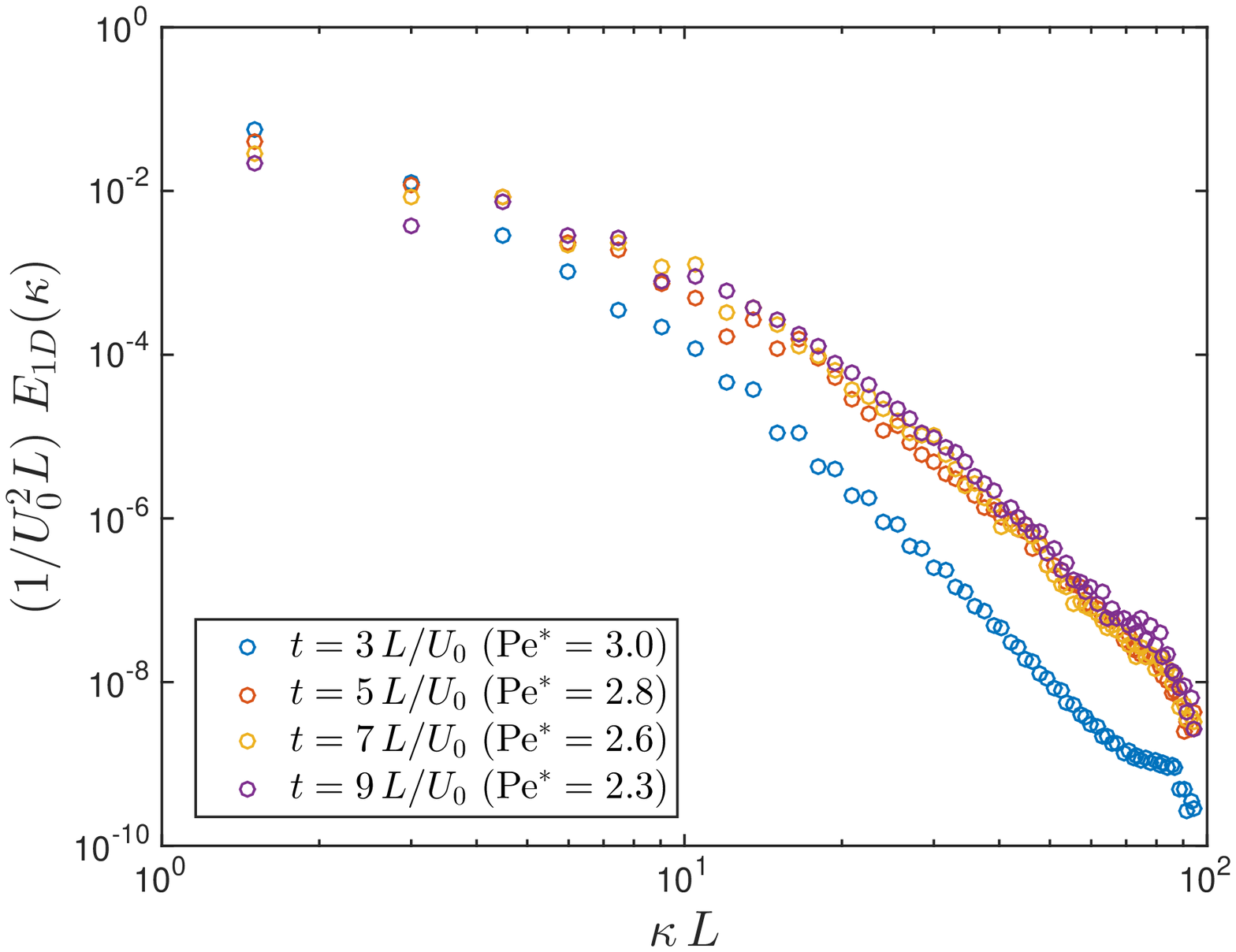}
\hfill \includegraphics[width=0.45\textwidth]{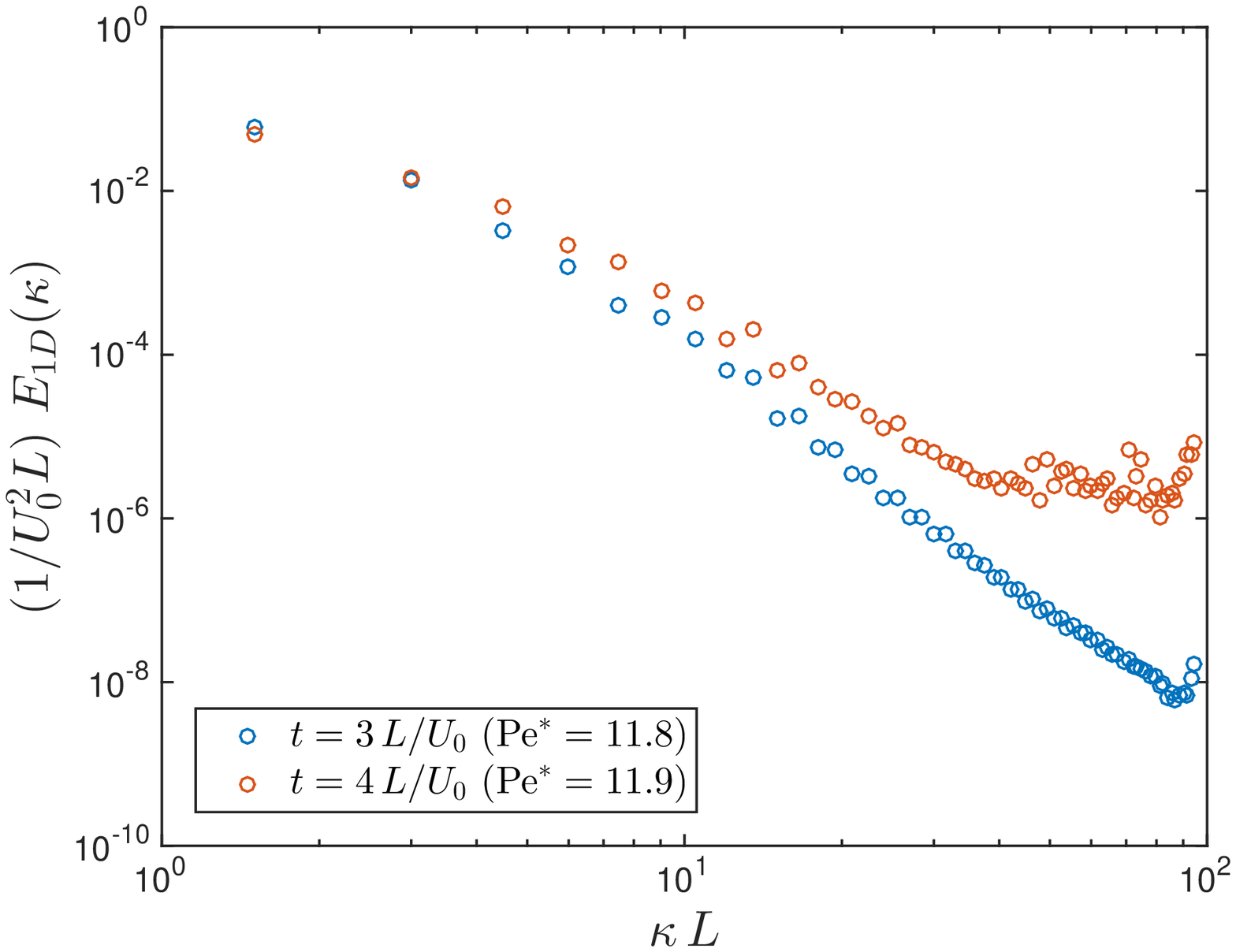}
\hfill \includegraphics[width=0.45\textwidth]{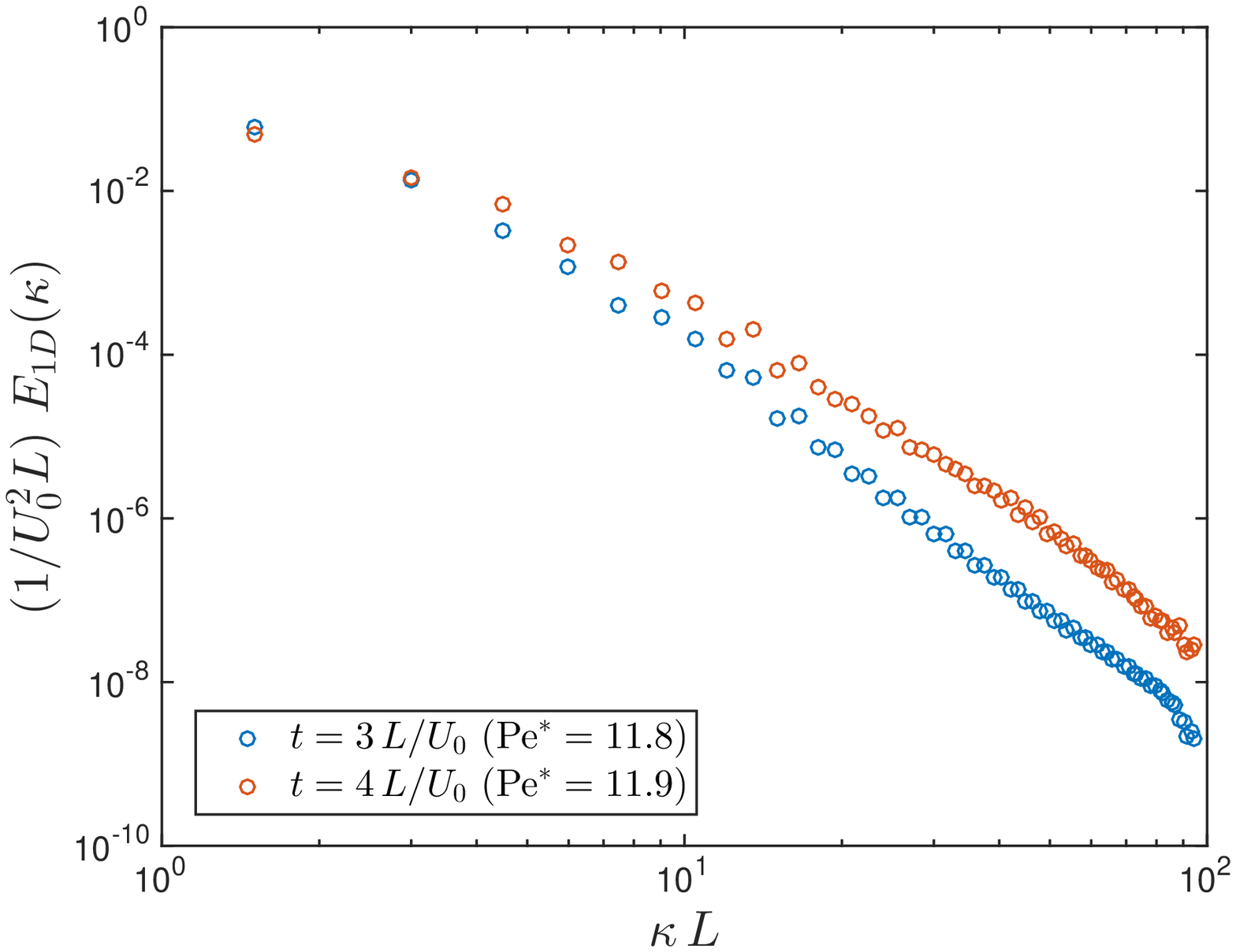}
\hfill \includegraphics[width=0.45\textwidth]{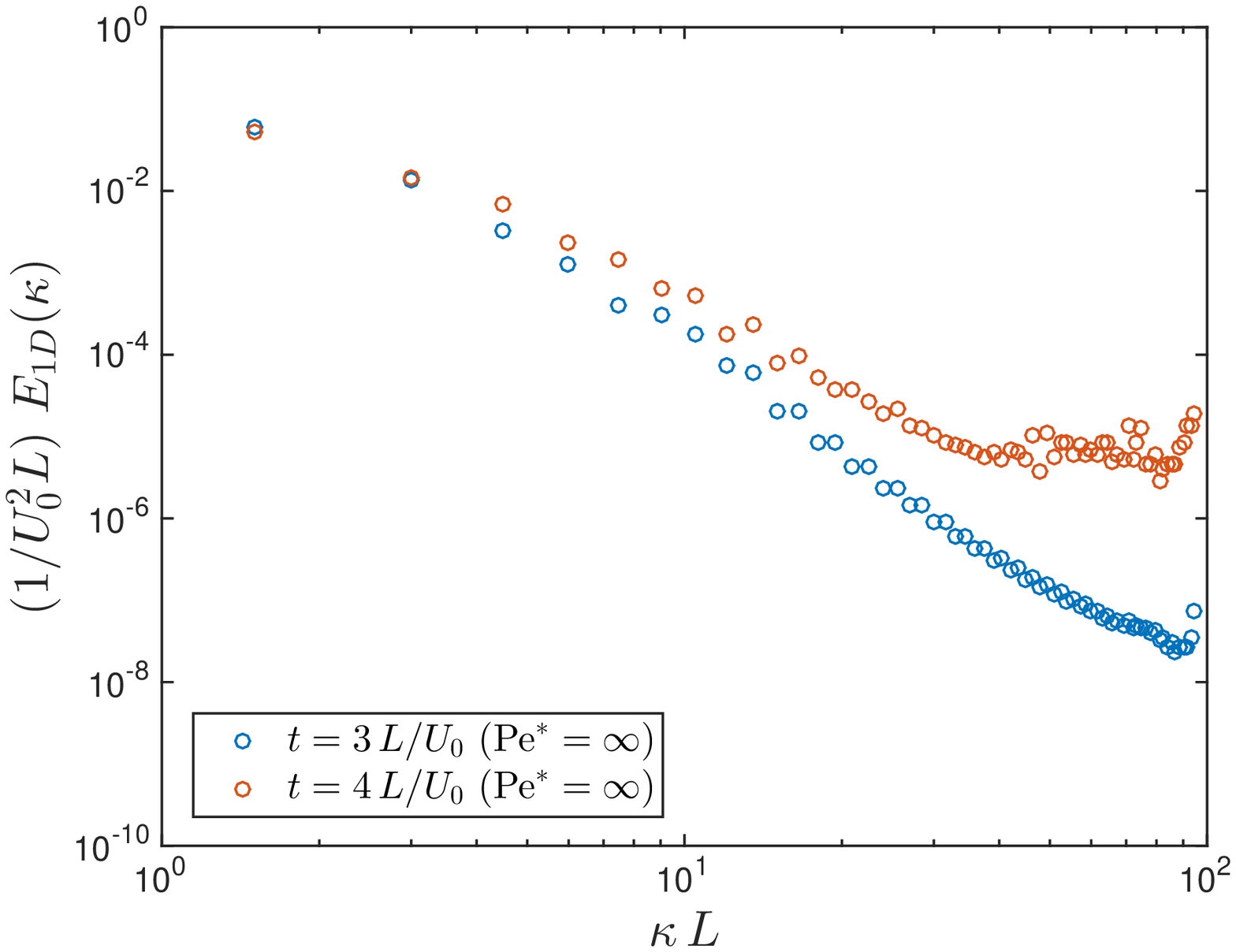}
\hfill \includegraphics[width=0.45\textwidth]{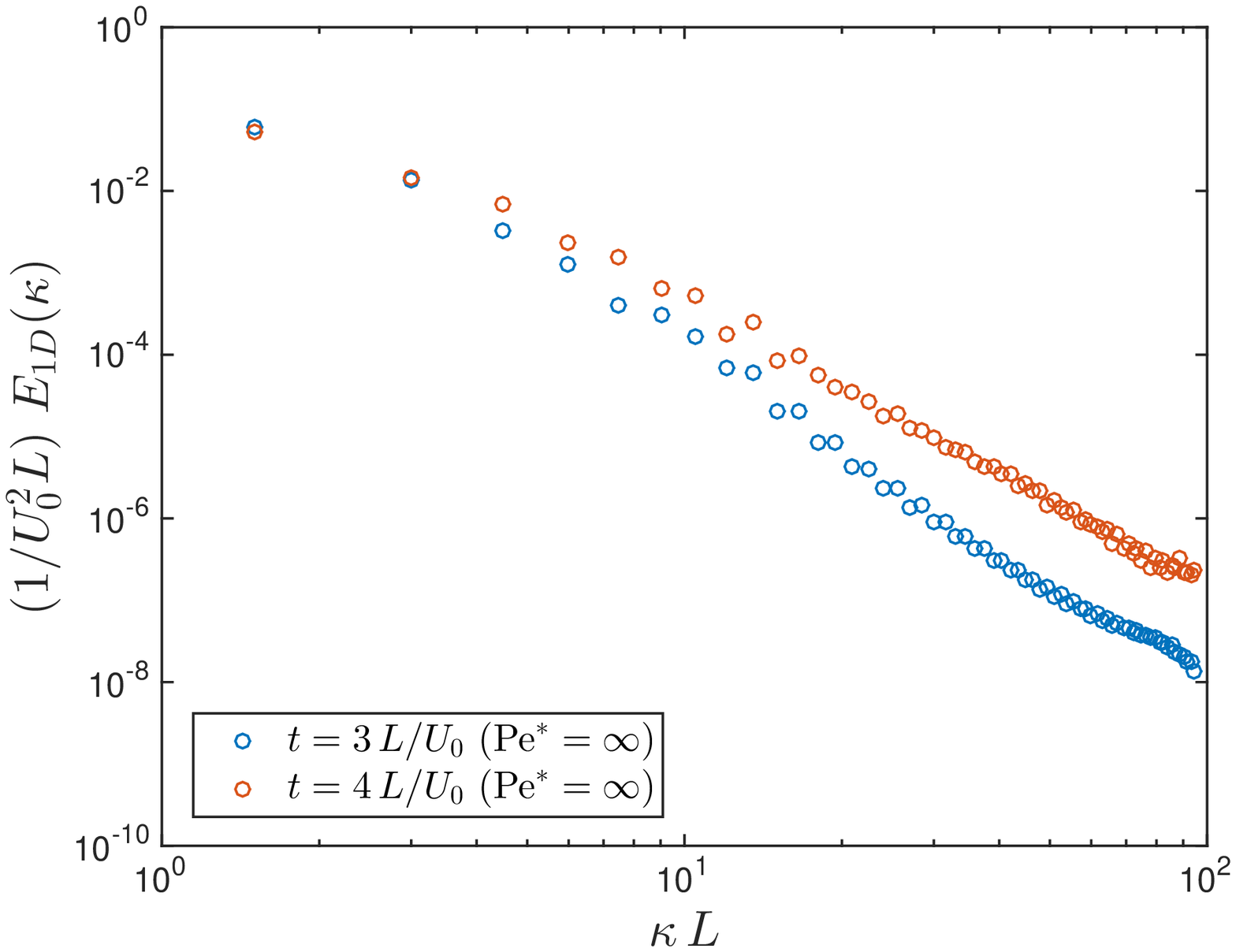}
\caption{\label{f:kinEnSpTGV} Time evolution of one-dimensional kinetic energy spectra for the Taylor-Green vortex at (from top to bottom) $\textnormal{Re} = 100$, $400$, $1600$ and $\infty$ with $\beta = 0.25$ (left) and $\beta = 1.00$ (right). The modified P\'eclet number $\textnormal{Pe}^*$ at each time is indicated in the legend. The right limit of the $x$ axis corresponds to the Nyquist wavenumber of the grid.}
\end{figure}
\begin{figure}
\centering
\includegraphics[width=0.45\textwidth]{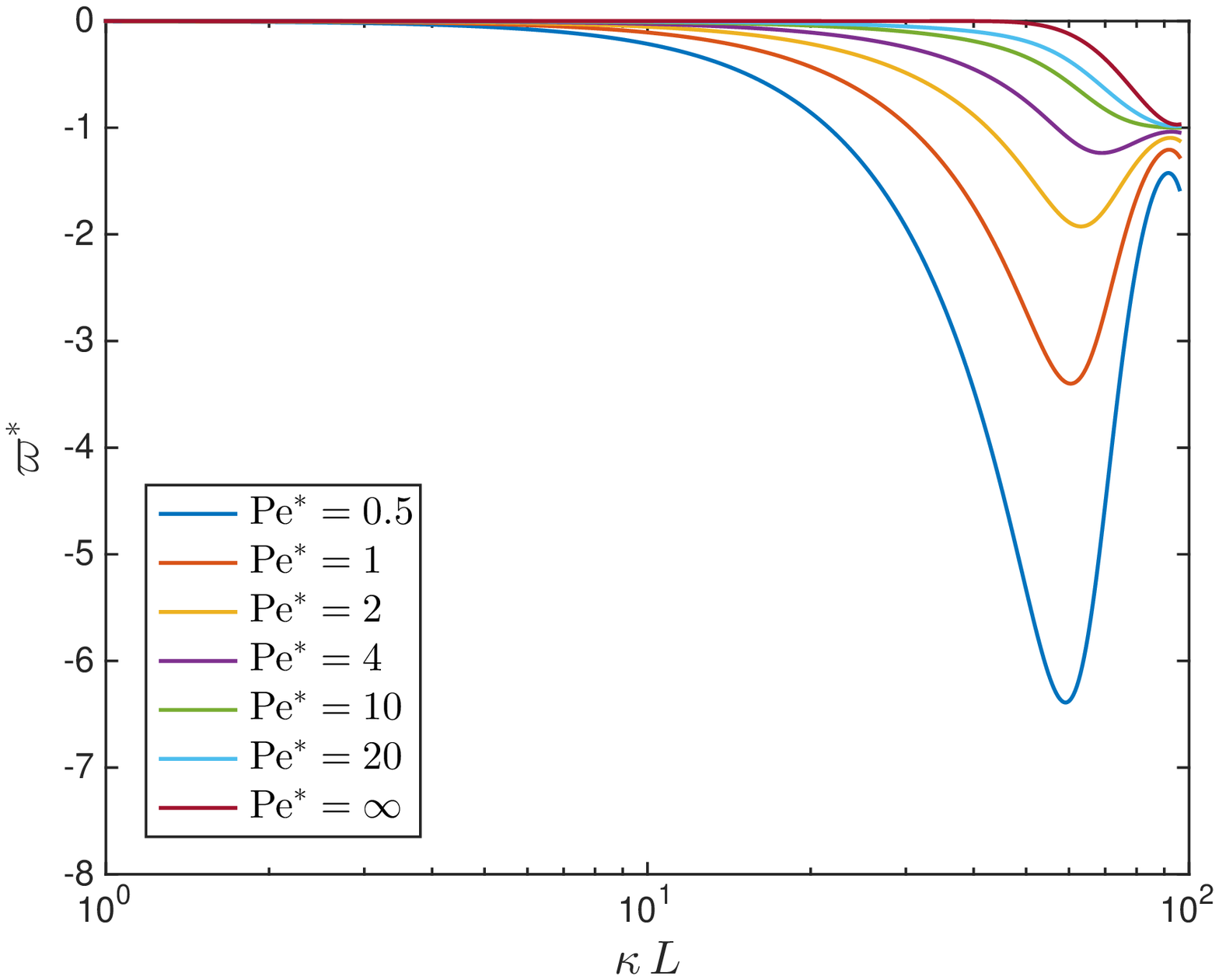}
\hfill \includegraphics[width=0.45\textwidth]{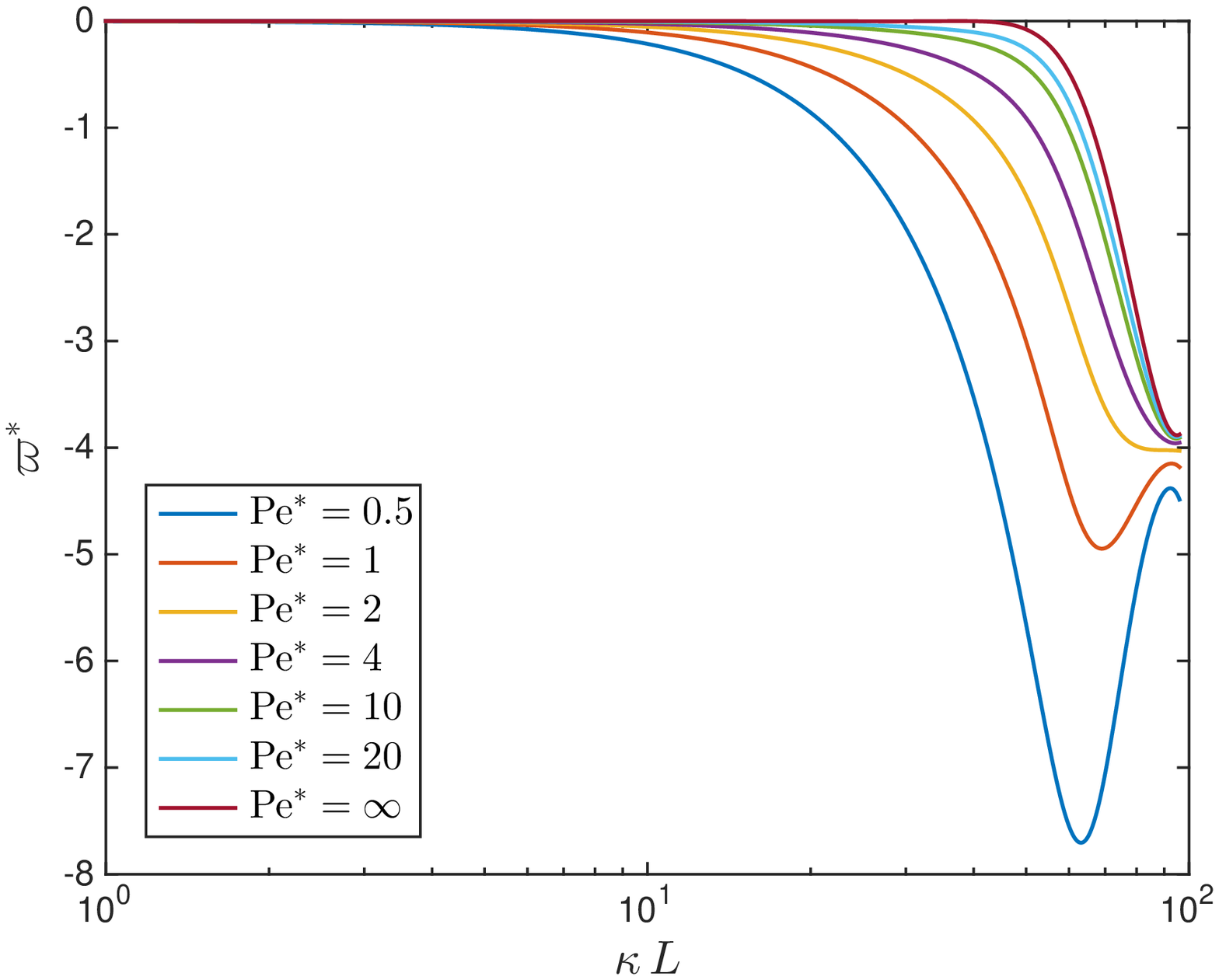}
\caption{\label{f:shortTimeDiffusionTGV} Short-term diffusion curves with $\beta = 0.25$ (left) and $\beta = 1.00$ (right) from non-modal analysis. These curves are to be compared to the energy spectra for the Taylor-Green vortex in Figure \ref{f:kinEnSpTGV}. The right limit of the $x$ axis corresponds to the Nyquist wavenumber of the grid.}
\end{figure}

Figure \ref{f:kinEnSpTGV} shows the time evolution of one-dimensional kinetic energy spectrum at the Reynolds numbers considered. The left and right images in these figures correspond to $\beta = 0.25$ and $1.00$, respectively, and the value of $\textnormal{Pe}^*$ at each time is indicated in the legends. The modified P\'eclet number in this problem is defined as $\textnormal{Pe}^* = h^* \, \rho u|_{rms} / \mu$, where $\rho u|_{rms}$ is the root mean square momentum. We note that, for a given $\textnormal{Re}$, the P\'eclet number slightly changes over time due to differences in $\rho u|_{rms}$. The short-term diffusion curves from non-modal analysis at the relevant P\'eclet numbers are shown in Figure \ref{f:shortTimeDiffusionTGV}, where the $x$ axis has been mapped from $\kappa \, h / (P+1)$ (as in Figures \ref{f:shortTimeDiffusionCurves_pStudy}$-$\ref{f:shortTimeDiffusionCurves_betaStudy}) to $\kappa \, L$ to facilitate the comparison with the energy spectra.

Like in Section \ref{s:burgers}, non-modal analysis results show good agreement with the turbulent energy spectrum in the simulations. First, energy pileups at large wavenumbers are observed in the spectrum when the short-term diffusion curves are non-monotonic and diffusion decreases after a maximum. Particularly informative is the Reynolds number $400$. From non-modal analysis, the short-term diffusion curves at the corresponding $\textnormal{Pe}^* \approx 2.0 - 3.0$ are non-monotonic and monotonic near the Nyquist wavenumber with $\beta = 0.25$ and $1.00$, respectively. As a consequence, energy accumulates at large wavenumbers with $\beta = 0.25$; which does not occur with standard upwinding.

Second, non-modal analysis predicts a small diffusion at high wavenumbers with under-upwinding in convection-dominated regimes, and this directly translates to the TGV results. In particular, when the physical viscosity is small (i.e.\ in the high Reynolds number cases), the dissipation at high wavenumbers with $\beta = 0.25$ does not suffice to dissipate all the energy that is being transferred from the larger scales through the turbulence cascade. As a consequence, and despite the diffusion curves are monotonic, energy starts to accumulate near the Nyquist wavenumber from the beginning of the simulation. As time evolves, this accumulation extends to larger scales due to the insufficient dissipation of energy at high wavenumbers, and eventually leads to nonlinear instability and the simulation breakdown at times $t \approx 4.42 \, L / U_0$ and $4.01 \, L / U_0$ for $\textnormal{Re} = 1600$ and $\infty$, respectively. In addition to non-monotonic dissipation characteristics, insufficient dissipation (specially at large wavenumbers) is \textit{per se} another mechanism for nonlinear instability in under-resolved turbulence simulations.

\section{\label{s:discussion}Guidelines for the construction of superior schemes}

From the non-modal analysis and the numerical experiments presented in this paper, we can deduce essential guidelines to improve the accuracy and numerical robustness of under-resolved simulations of nonlinear systems of equations. In particular, to construct robust under-resolved simulations of nonlinear problems, especially featuring a kinetic energy cascade, such as turbulent flows, the numerical scheme should possess 
\begin{itemize}
\item monotonic and slowly varying short-term diffusion characteristics, to avoid energy bottlenecks at some specific wavenumbers;
\item sufficient dissipation near the Nyquist wavenumber of the grid, to avoid energy accumulation at large wavenumbers. 
\end{itemize}
If these two requirements are not met, the numerical computation can develop nonlinear instabilities that will ultimately result in the crash of the simulation.

On the other hand, if these two requirements are met, the numerical scheme will resemble a built-in `subgrid-scale model' for under-resolved turbulence simulations, that can be directly comparable to explicit filtering and subgrid-scale modeling techniques adopted in the context of LES, including variational multiscale \cite{Collis:2002,Hughes:1998,Hughes:2000,Murman:2014} and spectral vanishing viscosity \cite{Karamanos:2000,Kirby:2002,Tadmor:1989} methods. 
In particular, non-modal analysis indicates that, for moderately high accuracy orders, hybridized DG methods introduce numerical dissipation in under-resolved computations of convection-dominated flows, and this dissipation is localized near the Nyquist wavenumber. This implicit subgrid-scale model indeed resembles variational multiscale and spectral vanishing viscosity approaches in the sense that dissipation is applied to the smallest resolved scales and the amount of such dissipation depends mostly on the energy in those scales. Therefore, by choosing the element size $h$ and the polynomial order $P$ inside of each element, the diffusion properties of the scheme can be tuned to obtain an equivalent filter width that can be readily used within an implicit LES context. The information regarding diffusion properties can also be used to improve accuracy in classical (explicit) LES, by better decoupling the wavenumber of the LES filter from the dissipation introduced by the numerics.

From a purely non-modal analysis standpoint, polynomial orders $P=2$, $3$ and $4$ with standard upwinding seem to be the most adequate for under-resolved simulations of turbulent flows, at least in the implicit LES context. For lower polynomial orders, dissipation is introduced in scales that are much larger than the grid resolution. Strong under/over-upwinding, as well as higher polynomial orders, may lead to stability issues. We note that Riemann solvers that are based on the maximum-magnitude eigenvalue of the Jacobian matrix of the Euler fluxes, such as the Lax-Friedrichs and the HLL solvers \cite{Toro:1999}, produce over-upwinding at low Mach numbers.

\section{\label{s:conclusions}Conclusions}

We introduced a non-modal analysis framework to investigate the short-term dynamics, and in particular the short-term diffusion in wavenumber space, of the semi-discrete system arising from the spatial discretization of the linear convection-diffusion equation. While applicable to high-order SEM in general, the non-modal analysis methodology was illustrated for the particular instance of the hybridized DG methods. The effect of the polynomial order, the P\'eclet number and the upwinding parameter on the short-term diffusion of the hybridized DG scheme were investigated. From these studies, the diffusion characteristics that are better suited, in terms of accuracy per DOF and robustness, for under-resolved simulations of nonlinear systems, seem to be those for polynomial orders $P=2$, $3$ and, to a lessen extent, $4$. Beyond these polynomial orders, the diffusion curves become strongly non-monotonic; which may lead to nonlinear instability due to bottlenecks in the energy spectrum. Strong under-/over-upwinding, such as with Lax-Friedrichs type Riemann solvers at low Mach numbers, may similarly lead to stability and accuracy issues.

While devised in the linear setting, non-modal analysis succeeded to predict the trends observed in the nonlinear problems considered. In particular, non-modal analysis results showed excellent agreement with numerical results for the Burgers, Euler and Navier-Stokes equations. From a practical perspective, non-modal analysis provides insights on why high-order SEM may suffer from stability issues in under-resolved simulations, as well as on how to devise more robust schemes for these problems. Furthermore, it provides insights to understand and improve the built-in subgrid-scale model in the scheme for under-resolved turbulence simulations. A discussion on these topics was presented.

The non-modal analysis framework can be generalized in several ways. First, one may study the finite-time behavior of the system (for some fixed $t > 0$), as opposed to its short-term behavior (limit $t \downarrow 0$). Second, the analysis can be extended to arbitrary initial conditions, instead of Fourier modes only. We note, however, that Fourier modes are arguably the best choice to provide insights on the robustness and accuracy of the scheme. Third, more complex discretizations, including non-uniform meshes, non-constant coefficients and multi-dimensional problems, could be considered. These generalizations, however, would add more parameters to the analysis and partially defeat its purpose; which is to provide with a tool that, with a few inputs, approximately describes the behavior of the scheme for nonlinear problems.

\section*{Acknowledgments}

The authors acknowledge the Air Force Office of Scientific Research (FA9550-16-1-0214) and Pratt \& Whitney for supporting this effort. The first author also acknowledges the financial support from the Zakhartchenko Fellowship.

\appendix

\section{\label{s:connectionStHyDG}Connection between standard DG and hybridized DG for linear convection}

\subsection*{Standard DG}

We consider standard DG numerical fluxes of the form
\begin{subequations}
\label{e:fluxDG}
\begin{alignat}{2}
 \widehat{f}_{h,\ominus} = a \, \frac{u_{h,\ominus} + u^L_{h,\oplus}}{2} - \beta \, |a| \, \frac{u_{h,\ominus} - u^L_{h,\oplus}}{2} \mbox{ ,} \\
 \widehat{f}_{h,\oplus} = a \, \frac{u_{h,\oplus} + u^R_{h,\ominus}}{2} + \beta \, |a| \, \frac{u_{h,\oplus} - u^R_{h,\ominus}}{2} \mbox{ ,}
 \end{alignat}
 \end{subequations}
where $\beta \geq 0$ is the upwinding parameter. The cases $\beta = 0$ and $\beta = 1$ correspond to the central flux and the standard upwinding, respectively.

\subsection*{Hybridized DG}

We consider hybridized DG numerical fluxes of the form
\begin{subequations}
\label{e:fluxhDG}
\begin{alignat}{2}
\tag{\ref{eq:fluxmi}}
\widehat{f}_{h,\ominus} = a \, \widehat{u}_{h,\ominus} - \beta \, |a| \, (u_{h,\ominus} - \widehat{u}_{h,\ominus}) \mbox{ ,} \\
\tag{\ref{eq:fluxpl}}
\widehat{f}_{h,\oplus} = a \, \widehat{u}_{h,\oplus} + \beta \, |a| \, (u_{h,\oplus} - \widehat{u}_{h,\oplus}) \mbox{ .}
\end{alignat}
\end{subequations}
For $\beta > 0$, the numerical trace is uniquely defined and given by $\widehat{u}_h = (u^L_{h,\oplus} + u^R_{h,\ominus}) / 2$. The standard DG numerical fluxes \eqref{e:fluxDG} are therefore recovered, and both the standard and hybridized DG schemes lead to the same numerical solution $u_h$. We note that the hybridized DG discretization is singular for $\beta = 0$, and in particular
\begin{equation}
\left( \begin{array}{c}
\delta \tilde{u}_h^L \\
\delta \widehat{u}_h \\
\delta \tilde{u}_h^R
\end{array}
\right) = \left( \begin{array}{c}
D^{-1} \tilde{\phi}_\oplus \\
1 \\
- D^{-1} \tilde{\phi}_\ominus
\end{array}
\right) 
\end{equation}
is in the nullspace.

\section*{References}


\end{document}